\documentclass[pre,aps,showpacs,byrevtex,onecolumn]{revtex4}
\usepackage{amssymb}

\usepackage{graphicx}
\usepackage{amsmath}

\begin{document}

\title{Equation of state of looped DNA}
\author{Igor M. Kuli\'{c}$^1$, Herv\'{e} Mohrbach$^2$, Rochish Thaokar$^3$, and
Helmut Schiessel$^2$}
\date{\today}

\address{$^1$Department of Physics and
Astronomy, University of Pennsylvania Philadelphia, PA 19104,
USA\\
$^2$Instituut-Lorentz, Universiteit
Leiden, Postbus 9506, 2300 RA Leiden, The Netherlands\\
$^3$Department of Chemical Engineering, Indian Institute of
Technology, Bombay, Mumbai, 400076, India}

\begin{abstract}
We calculate the equation of state of DNA under tension for the case that
the DNA features loops. Such loops occur transiently during DNA condensation
in the presence of multivalent ions or sliding cationic protein linkers. The
force-extension relation of such looped DNA modelled as a wormlike chain is
calculated via path integration in the semiclassical limit. This allows us
to determine rigorously the high stretching asymptotics. Notably the
functional form of the force-extension curve resembles that of straight DNA,
yet with a strongly renormalized apparent persistence length. That means
that the experimentally extracted single molecule elasticity does not
necessarily reflect the bare DNA stiffness only, but can also contain
additional contributions that depend on the overall chain conformation and
length.
\end{abstract}


\maketitle

\section{Introduction}

The DNA double helix is the molecule that encodes the genetic information in
living cells. In addition to carrying the genome, DNA has also specific
physical properties that are essential for its biological functions. Its
mechanical properties are exploited by the protein machinery for the
transcription, replication, repair and packaging of DNA \cite{Alberts}.
During the last decade it has been possible to manipulate single DNA
molecules to determine its elastic properties under different physical
conditions \cite{Strick}. In these experiments, the extension of single
molecule versus an applied stretching force is measured by a variety of
technics including magnetic beads \cite{Smith,Strick1}, optical traps \cite
{Smith2,Wang}, micro-needles \cite{Cluzel}, hydrodynamic flow \cite{Perkins}
and AFM \cite{Rief}. The studies also made it possible to better understand
mechanical interactions between DNA and proteins \cite{Bustamante}.

The most appealing theoretical description of the DNA molecule is the
wormlike chain (WLC) model which is a coarse-grained model of DNA with a
single parameter, the persistence length $l_p$, characterizing the chain
stiffness. Originating back to the first half of the last century~\cite
{Frenkel-Kratky-Porod} it gained renewed interest after the semiflexible
nature of DNA and other (bio)polymers became clear \cite{WLCExp} and it is
now indispensable for the theoretical understanding of many single molecule
experiments. In many cases stiff polymers show a characteristic
force-extension behavior that can be well understood in terms of the WLC as
being the result of entropic fluctuations of the chain that -- with
increasing tension -- become suppressed at shorter and shorter wave lengths.
Measuring the force-extension characteristics of such a chain allows to
extracts its overall contour length as well as its persistence length.

The DNA in living cells is rarely found in its straight ''naked'' state;
rather an overwhelming fraction of DNA is strongly configurationally
constrained by binding proteins causing loops, bends and wraps. In
particular, protein complexes forming loops are essential for biological
process like distant gene expression or DNA packaging \cite{Alberts}. The
formation of loops in a DNA molecule under tension has been the subject of
the theoretical investigation in Ref.~\cite{Markoloop}. Also, single
molecule stretching experiments on DNA condensed with multivalent
counterions performed by several groups \cite{toroids,toroids2} might bear
loops or related structures like DNA toroids \cite{RacquetsRodsToroids}.
While the statistical mechanics of unconstrained DNA under tension is well
described by the WLC \cite{WLCExp}, the presence of topological constraints
like supercoiling \cite{Strick2,BouchMezard} and entanglements \cite{Kardar}%
, or geometrical constraints like protein induced kinks and bends \cite
{BruinsmaRudnick,Marko} renders analytical results more difficult.

\begin{figure}[ptb]
\includegraphics*[width=8cm]{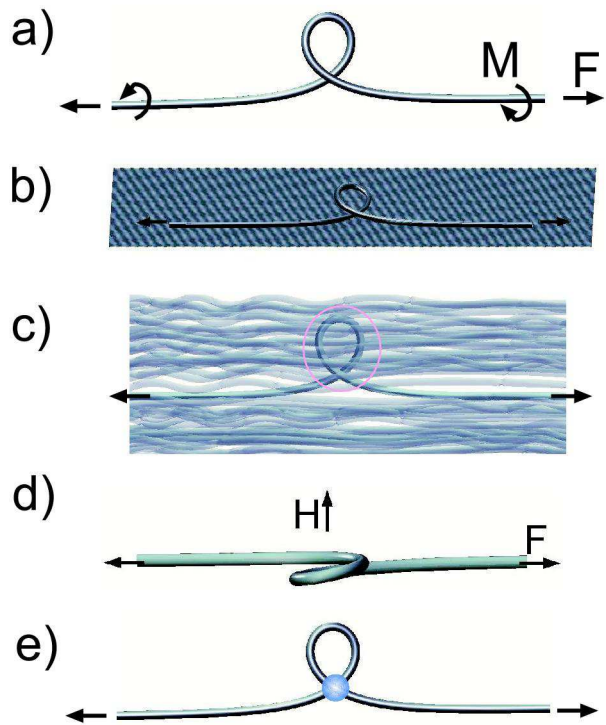}
\caption{Various examples of stable loops in DNA under tension: (a) Applied
torque M at the ends. (b) DNA adsorbed on a surface. (c) DNA surrounded by a
dense solution of infinitely long DNAs. Unfolding of the loop goes hand in
hand with an energetically costly transient ''cavity'' creation (in the pink
region). (d) DNA in a strong magnetic field H perpendicular to the applied
force. (e) DNA looped by a freely sliding linker ligand (''weakly
condensed'' DNA).}
\label{LoopCompilation}
\end{figure}

In this paper, we expand the repertoire of analytically solvable
''equations of state'' by deriving the force extension relation
for a DNA with a sliding loop as depicted in
Fig.~\ref{LoopCompilation}. The computation is performed by
evaluating quadratic fluctuations around the looped solution -- a
non-constant saddle-point of the DNA elastic energy. The method is
essentially analogous to the semiclassical treatment of tunneling
amplitudes in quantum mechanics and instantons in quantum field
theory \cite {Pathintegral}. The equation of state of looped DNA
that we present here can be considered as a paradigmatic model
case for stretching DNA with a non-trivial ground state. Having
understood the physics of the looped DNA it is straightforward to
extend the analysis also to other cases where the overall DNA
conformation is far from being straight. The calculation presented
in the present work has been sketched in a previous paper \cite
{kulic} (for the high-force limit) together with some interesting
experimental situations, namely rigid protein-induced kinks and
anchoring deflections in AFM stretching of semiflexible polymers.
Expressions relating the force-extension measurements to the
underlying kink/boundary deflection geometry were also provided in
Ref.~\cite{kulic} and applied to the case of the GalR-loop complex
\cite{GalR}.

It is not the purpose of this paper to analyze concrete experimental setups,
but to give a more detailed and comprehensive description of our
computation. This paper is also more general as we employ our method also to
determine the equation of state of sliding loop in the small force regime
where the semi-classical approximation is valid for DNA with length smaller
than the persistence length whereas for longer DNA our computation is only
valid for large forces. In that strong force regime, it will be shown that
the presence of the loop modifies the the elastic response of the chain in
such a manner that the persistence length appears effectively reduced
according to the relation
\begin{equation}
l_{P}^{app}=l_{p}\left( 1+8\frac{l_{p}}{L}\right) ^{-2}  \label{Intro1}
\end{equation}
For a single loop this is obviously a finite size effect involving the
scaled total length $l_{p}/L$, but the effect remains significant over a
large range of parameters (e.g. $l_{P}^{app}\approx0.74l_{P}$ for $%
L/l_{p}=50 $). Therefore the interpretation of corresponding stretching data
has to be taken with care: Even though the data seem to suggest WLC
behavior, the extracted value of persistence length might not reflect the
real chain stiffness.

An intriguing example that actually inspired this work is the
force-extension characteristics of DNA in the presence of condensing agents
like spermidine or CoHex \cite{toroids}. It shows in some cases a
stick-release pattern which might be attributed to the sequential unpeeling
of single turns of a toroidal condensate \cite{kulic04}. What is important
here is that in between the force peaks one can nicely fit WLC behavior but
the persistence length that one extracts from these data is typically much
lower than that of DNA. Only when the last turn is disrupted and the DNA is
in a straight configuration one finds the expected value of the chain
stiffness.

Before going into any kind of theoretical analysis of looped DNA under
tension it is important to note that such a configuration is intrinsically
\textit{not} stable and has therefore to be stabilized by some mechanism.
Some possible mechanisms are listed in Fig.~\ref{LoopCompilation}: (a)
supercoiling in twisted DNA (the same phenomenon like in a looped telephone
cable), (b) DNA adsorption on a surface (e.g.~a liquid membrane) \cite
{maier99}, (c) DNA in a dense liquid crystalline environment kinetically
prohibiting the loop unfolding, (d) DNA in a strong magnetic field that
tends to align it in a plane perpendicular to the field lines \cite{Mareth}
and (e) DNA condensed by multivalent counterions and other ligands that form
a freely sliding link.

In Section II of this paper we will shortly review the Euler-Kirchhoff
elastic description of the (constrained) ground states of DNA under tension.
It is extremely useful for understanding the behavior of constrained ''cold
DNA''. By ''cold DNA'' we metaphorically mean DNA in situations where the
importance of its configurational entropy is negligible as compared to its
elastic energy. This is typically the case for short DNA lengths (below its
persistence length $l_{P}$) and large energy densities (larger than tens of $%
k_{B}T$'s per $l_{P}$). In the second subsection we switch on the
temperature and discuss how the thermal DNA wiggling affects its
behavior. As mentioned above such ''hot DNA'' responds purely
entropically to moderate pulling forces. We review the well known
derivation of its mechanical ''equation of state'', i.e., the
force extension behavior of stretched DNA. In Section III we
derive the statistical mechanics for looped DNA under tension for
the most simple case where the looped DNA is confined to two
dimensions as depicted in Fig.~\ref{LoopCompilation}b. In this
context we will learn how stretched DNA behaves when its new
''ground state'' is far from the straight configuration. The
analytical machinery that is applied and developed further here
has its roots in classical problems of physics like instantons in
quantum mechanical tunnelling \cite{Pathintegral}. The unifying
concept behind all these phenomena is that of path integration in
the semiclassical limit. In Section IV we finally calculate the
stretching of looped DNA in three dimensions. We start with the
case of DNA being oriented in a strong magnetic field,
Fig.~\ref{LoopCompilation}d. After having given a rigorous
derivation of this case, we determine the partition function for a
loop stabilized by a sliding ligand, cf.~Fig.~\ref
{LoopCompilation}e which finally leads to Eq.~\ref{Intro1}.

\section{Introductory remarks}

\subsection{Euler elastica}

The basic assumption of a purely elastic description of DNA (and other
semiflexible polymers as well) is that the local energy density of a given
DNA state is given as a quadratic function of the underlying distortions
from the straight state. Let us consider the simplest situation where the
DNA twist degree of freedom can be neglected. This can be done in cases when
the DNA twist is not constrained from outside, i.e., when no external
torsional torques are acting on it. Then we can describe the path of the DNA
of given length $L$ and bending constant $A$ subjected to an applied tension
$\mathbf{F}$ by the space curve $\mathbf{r}\left( s\right) $\ with the
tangent $\mathbf{t}\left( s\right) =\frac{d}{ds}\mathbf{r}\left( s\right) $.
It is convenient to choose the parameter $0<s<L$ as the contour length and
to normalize the tangent to unity $\mathbf{t}^{2}\left( s\right) =1$. The
elastic energy under an applied force $F$ writes in this case~\cite
{Landau-Lifschits}
\begin{equation}
E\left[ t\right] =\int\nolimits_{0}^{L}\left( \frac{A}{2}\left( \frac{d%
\mathbf{t}}{ds}\right) ^{2}-\mathbf{F\cdot t}\right) ds  \label{EEElastic}
\end{equation}
Here $A$ is the bending stiffness that is usually expressed as $%
A=l_{P}k_{B}T $ where $l_{P}$ is the orientational persistence length; for
DNA $l_{P}\approx50$ nm \cite{hagerman}.

Let us look first at ''cold'' DNA, i.e., at a molecule shorter than $l_{P}$
where we can in principle neglect entropic contributions to its behavior.
The problem of finding the DNA conformation reduces in this case to the
classical problem of inextensible elastic beam theory~\cite{Landau-Lifschits}
of finding the energy minimizing state $\delta E/\delta\mathbf{t}=0$ which
satisfies the given constraints. In a concrete computation one would
parametrize the unit tangent vector $\mathbf{t}$ in spherical coordinates $%
\mathbf{t}=\left( \cos\phi \sin\theta,\sin\phi\sin\theta,\cos\theta\right) $
and put the force along the $z-$axis so that the energy now writes
\begin{equation}
E=\int_{0}^{L}\left[ \frac{A}{2}\left( \dot{\phi}^{2}\sin^{2}\theta +\dot{%
\theta}^{2}\right) -F\cos\theta\right] ds  \label{EbendEulers}
\end{equation}
Note that this linear elastic ansatz can be readily extended to the
description of twisted DNA states \cite{Benham} by introducing another
degree of freedom, the twisting angle $\psi\left( s\right) $, in addition to
$\phi\left( s\right) $ and $\theta\left( s\right) $. In this case one has to
modify Eq.~\ref{EbendEulers} by adding the term $\frac{B}{2}\left( \dot{\phi}%
\cos\theta+\dot{\psi}\right) ^{2}$ with $B$ denoting the twist-rigidity
constant that is for DNA of the same magnitude as the bending constant, $%
B\approx70k_{B}T$ nm \cite{hagerman}. The reason why we can neglect it in
some (but by far not all) problems is that if the twist angle $\psi$ is not
explicitly constrained (no rotational torque or torsional constraining of
DNA) $\psi$ can always adapt so that the $B$ multiplying term in the
integral vanishes (without affecting $\phi\left( s\right) $ and $%
\theta\left( s\right) $).

Remarkably, as pointed out by Kirchhoff~\cite{Kirchhoff} the \textit{total
energy} of deformed DNA (elastic rod) can be mapped onto the \textit{%
Lagrangian action} of a symmetric spinning top in a gravity field. The
angles then $\theta\left( s\right) ,\phi\left( s\right) $ and $\psi\left(
s\right) $ describing the local deformations of the rod along the \textit{%
contour length} $s$ become the Euler-angles $\theta\left( t\right)
,\phi\left( t\right) $ and $\psi\left( t\right) $ of the spinning top
describing the rotation of the internal coordinates system (with respect to
the space fixed frame) as functions of \textit{time} $t$. All the quantities
appearing in Eq.~\ref{EbendEulers} have their counterparts in the spinning
top case~\cite{Nizette/Goriely}. The tension $F$ is the equivalent of the
gravity force acting on the spinning top; the rigidity constants $B$ and $A$
correspond to the principal moments of inertia around the symmetry axis and
perpendicular to it, respectively. The resulting rod shapes are usually
called \textit{Euler-Kirchhoff filaments} (in 3D) or \textit{Euler-elastica}
(in the 2D case). They are given explicitly in terms of elliptic functions
and integrals~\cite{Nizette/Goriely}. The latter fact allows one in many
cases (for a given set of forces and boundary conditions) to obtain the DNA
shapes in an analytical or at least numerically inexpensive manner.

\subsection{Semi-classical straight DNA stretching}

The previous description of DNA conformations via the ground state of a
purely elastic beam can, however, only be successful for very short DNA
(shorter than its persistence length $l_{P}$). In many practical situations
with the DNA molecules having lengths on the order of microns to centimeters
($\gg l_{P}$) one needs to go beyond the ground state description. But as
shown in this paper, this is even true for short DNA ($<l_{P}$) forming a
loop. An important question (from the experimental and theoretical point of
view) in the context of DNA stretching is the determination of the mean
end-to-end distance of a DNA chain as a function of the stretching force $F$
at a finite temperature $T$ \cite{WLCExp}. Then in order to take into
account temperature effects on a WLC under tension one has to compute the
following partition function
\begin{equation}
Q(\mathbf{F,}L,T)=\int \delta \left( \mathbf{\underline{t}}^{2}-1\right)
\mathcal{D}^{3}\left[ \mathbf{\underline{t}}\right] e^{-\beta E\left[
\mathbf{\underline{t}}\right] }  \label{PathIntegral1}
\end{equation}
with the energy expression given by Eq.~\ref{EEElastic}. This partition
function is formally the imaginary time analytical continuation of a path
integral of a quantum particle on a unit sphere subjected to an external
force. The chain inextensibility constraint (represented by the $\delta $
function in Eq.~\ref{PathIntegral1}) makes this path-integral a highly
non-trivial quantity to compute as it introduces a parametrization-dependent
nontrivial measure term. But as shown below by choosing a proper
parametrization of the unit-sphere this unpleasant term does not give any
contribution if we limit the computation to the semiclassical approximation.

Let us briefly rederive the well-known results for the force-extension
behavior~\cite{WLCExp}. To evaluate the path integral we introduce the
following representation of the tangent vector defined in a Cartesian
coordinate system \cite{Pathintegral} $\mathbf{\underline{t}}=(\sqrt{1-q^{2}}%
,q_{y},q_{z})$. If we now consider a force that points in the $x$ direction,
then the energy becomes only $q$ dependent and reads
\begin{equation}
E\left[ q\right] =\int\nolimits_{0}^{L}\left( \frac{A}{2}g_{ij}\left(
q\right) \frac{dq^{i}}{ds}\frac{dq^{j}}{ds}+F\sqrt{1-q^{2}}\right) ds
\label{xaxis}
\end{equation}
with $\left( i,j=y,z\right) $. The metric and its determinant that determine
the $O(3)$ invariant measure are respectively $g_{ij}\left( q\right) =\delta
_{ij}+\frac{q_{i}q_{j}}{1-q^{2}}$ and $g\left( q\right) =\frac{1}{1-q^{2}}$,
and the partition function is now written in a curved space as
\begin{equation}
Q(F,L,T)=\int \mathcal{D}^{2}\left[ q\right] \sqrt{g\left( q\right) }%
e^{-\beta E\left[ q\right] }=\int \mathcal{D}^{2}\left[ q\right] e^{-\beta E%
\left[ q\right] -E_{m}\left[ q\right] }
\end{equation}
with a measure term that can be exponentiated $E_{m}\left[ \theta \right] =%
\frac{\delta \left( 0\right) }{2}\int\nolimits_{0}^{L}dt\log \left(
1-q^{2}\right) $ which is highly singular. To go further we now parameterize
the unit vector by the Euler angles $\phi (s)$ and $\theta (s)$. The choice
of the force direction parallel to the $x-$axis in Eq.~\ref{xaxis} turns out
to be necessary because an expansion of Eq.~\ref{EbendEulers} around the
straight configuration $\theta =0$ is singular as the angle $\phi (s)$ is
then arbitrary. This causes no technical problem when dealing with ground
states since $\phi $ enters Eq.~\ref{EbendEulers} only through its
derivative $\dot{\phi}$ but we need to rotate the force direction into the $%
x-$direction before dealing with the statistical mechanics of ''hot DNA'' on
basis of Eq.~\ref{PathIntegral1}. Introducing the angle $\vartheta
(s)=\theta (s)-\pi /2$ one has $\underline{\mathbf{t}}=\left( \cos \phi \cos
\vartheta ,\sin \phi \cos \vartheta ,\sin \vartheta \right) $. Having $%
J=\cos ^{2}\vartheta \cos \phi $ as the Jacobian of the transformation the
partition function writes:
\begin{equation}
Q(F\mathbf{,}L,T)=\int \mathcal{D}\left[ \phi \right] \mathcal{D}\left[
\vartheta \right] e^{-\beta E\left[ \vartheta ,\phi \right] -E_{m}\left[
\vartheta \right] }  \label{partitionfunction}
\end{equation}
with the elastic energy
\begin{equation}
E\left[ \vartheta ,\phi \right] =\int\nolimits_{0}^{L}\left( \frac{A}{2}%
\left( \dot{\phi}^{2}\cos ^{2}\vartheta +\dot{\vartheta}^{2}\right) +F\cos
\vartheta \cos \phi \right) ds  \label{elasticenergy}
\end{equation}
and with a measure expressed as
\begin{equation}
E_{m}\left[ \vartheta \right] =-\delta \left( 0\right)
\int\nolimits_{0}^{L}ds\log \left( {\left| \cos \vartheta \right| }\right)
\label{measure}
\end{equation}
which guarantees the $O(3)$ invariance of the measure $\mathcal{D}\left[
\phi \right] \mathcal{D}\left[ \vartheta \right] $ of the functional
integral. Our ''non-standard'' parametrization of the unit vector tangential
to the chain and the choice of a force pointing in the $x$-axis that looks
unusual are necessary in order to deal properly with the measure in a
semiclassical approach of the nontrivial functional Eq.~\ref
{partitionfunction}. Instead, the standard trick for WLC is based on an
analogy between the partition function and the Feynman amplitude of a
quantum particle. The partition function is then approximatively evaluated
by determining the eigenstates of the associated quantum Hamiltonian \cite
{WLCExp,BouchMezard}. This method seems to be difficult to adapt in the
presence of non-trivial saddle points, even though it has been applied for
tightly bend DNA \cite{Odijk2}.

Computationally Eq.~\ref{elasticenergy} with the two functions $\vartheta $
and $\phi $ entering the energy in a nonlinear manner makes the problem
difficult to be treated analytically. We therefore use the harmonic
approximation valid for small fluctuations around the straight configuration
($\mathbf{t}\parallel\mathbf{e}_{x}$), i.e., we expand the energy \ref
{elasticenergy} at the quadratic order around the trivial saddle point $%
\left( \phi_{0}=\vartheta _{0}=0\right) $. As $\beta$ is absent in front of
the measure, Eq.~\ref{measure}, this later does not participate to the
selection of the saddle point, but one has to take it into account when
considering quadratic fluctuations, i.e., $E_{m}\left[ \vartheta\right]
\approx-\delta\left( 0\right) \int\nolimits_{0}^{L}ds\vartheta_{0}^{2}$. The
saddle point being trivial the measure term vanishes and the partition
function factorizes into two independent partition functions:
\begin{equation}
Q(F,L,T)=e^{\beta FL}Q_{1}^{2}(F,L,T)
\end{equation}
with
\begin{equation}
Q_{1}(F,L,T)=\int\mathcal{D}\left[ \phi\right] e^{-\frac{\beta}{2}
\int\nolimits_{0}^{L}\left( A\dot{\phi}^{2}+F\phi^{2}\right) ds}
\label{harmonicpathintegral}
\end{equation}
Note that this factorization property and, in particular, the cancellation
of the measure are due to our choice of the coordinate system.

In order to compare later the free energy of the straight chain with that of
the looped configuration, we compute the path integral with the boundary
conditions $\phi(0)=\vartheta(0)=0$ and $\phi(L)=\vartheta(L)=0$ which are
the most convenient choice for a semi-classical evaluation of the path
integral around a non-trivial saddle point. The Fourier decomposition is
then restricted to sine functions $\phi(s)=\sum\sqrt{2/L}\sin\left( \omega
_{m}s\right) \phi_{m}$ with frequencies $\omega_{m}=\pi m/L$. The evaluation
of path integral then reduces to the computation of a product of Gaussian
integrals leading to
\begin{equation}
Q(F,L,T)=\frac{\beta\sqrt{AF}}{2\pi}\frac{e^{\beta FL}}{\sinh\left( L\sqrt{%
\frac{F}{A}}\right) }
\end{equation}
The force-extension relation can then be deduced from the expression $%
\left\langle \Delta x\right\rangle =-\partial G/\partial F$ where $G(F,L,T)=-%
\frac{1}{\beta}\text{ln}Q(F,L,T)$ is the free energy of the system. We then
obtain
\begin{equation}
\frac{\left\langle \Delta x\right\rangle }{L}=1+\frac{1}{2\beta FL}-\frac
{1}{2\beta\sqrt{FA}}\coth\left( L/\lambda\right)  \label{3dx1}
\end{equation}
Here we introduced the quantity $\lambda=\sqrt{A/F}$, usually called the
\textit{\ deflection length} or \textit{tension length} \cite
{OdijkQM,BruinsmaRudnick}, that becomes the relevant length scale in the
case of DNA under tension replacing the usual (tension-free) persistence
length $l_{P}=A/k_{B}T$.

>From the force-extension relation Eq.~\ref{3dx1} we see that two limiting
cases corresponding to regimes of small forces $L/\lambda\ll 1$ and large
forces $L/\lambda\gg1$ can be studied analytically.

\subsubsection{Small forces regime: $L/\protect\lambda\ll1$}

One can readily see that this condition implies a small force regime $\beta
FL\ll{l}_{p}{/L}$ which is compatible with the harmonic approximation only
if the persistence length is much larger than the chain length ($l_{p}\gg L$
). Then in this case Eq.~\ref{3dx1} becomes
\begin{align}
\frac{\left\langle \Delta x\right\rangle }{L} & \approx1-\frac{L}{6l_{p}}+
\frac{L^{3}}{24l_{p}}\frac{F}{A}\approx1-\frac{L}{6l_{p}}
\label{linearfluctu} \\
\end{align}
i.e. thermal fluctuations lead in leading order to a force-independent small
reduction of the end-to-end distance.

\subsubsection{Large forces regime $L/\protect\lambda\gg1$}

This regime implies the condition $\beta FL\gg l_{p}/L,$ that can be made
compatible with the harmonic approximation for any value of the ratio $%
l_{p}/L$. The free energy of the WLC under tension is then approximately
given by
\begin{equation}
G(F,L,T)\approx-FL+\frac{L}{\beta}\sqrt{\frac{F}{A}}-\frac{1}{\beta}\text{ln}%
\left( \frac{\beta\sqrt{AF}}{2\pi}\right)  \label{Glinear}
\end{equation}
whereas the force-extension relation in this limit becomes
\begin{equation}
\frac{\left\langle \Delta x\right\rangle }{L}\approx1-\frac{1}{2\beta\sqrt
{FA}}+\frac{1}{2\beta FL}  \label{3dxlargeF1}
\end{equation}
In this force regime the term $O(1/\beta FL)$ can always be neglected and
one arrives at the important formula~\cite{WLCExp}
\begin{equation}
\frac{\left\langle \Delta x\right\rangle }{L}\approx1-\frac{k_{B}T}{2\sqrt
{AF}}  \label{ExtensionForce}
\end{equation}
This can be solved for the force:
\begin{equation}
F\approx\frac{k_{B}T}{4l_{P}}\frac{1}{\left( 1-\left\langle \Delta
x\right\rangle /L\right) ^{2}}  \label{ForceExtensionWLC}
\end{equation}
The force, Eq.~\ref{ForceExtensionWLC}, is of entropic origin as the
proportionality to temperature indicates. Equation \ref{ForceExtensionWLC}
turned out to be a very powerful tool for directly and accurately
determining the persistence length of DNA molecules from micromanipulation
experiments under a multitude of conditions \cite{Strick}. One should note
that Eq.~\ref{ForceExtensionWLC} is only valid in the limit of large forces (%
$F\gg\frac{k_{B}T}{4l_{P}}=20$ fN) and large relative extensions $%
\left\langle \Delta z\right\rangle /L\approx O(1)$. Looking at its
simplicity it is somehow surprising that it is experimentally accurate for
piconewton forces almost up to the point where DNA starts to melt and the
WLC description breaks down (around 60 pN).

To have an expression that also includes very low forces (on the femtonewton
scale) one usually uses the following interpolation formula \cite{WLCExp}
\begin{equation}
F=\frac{k_{B}T}{l_{P}}\left[ \frac{1}{4\left( 1-\frac{\left\langle \Delta
z\right\rangle }{L}\right) ^{2}}-\frac{1}{4}+\frac{\left\langle \Delta
z\right\rangle }{L}\right]
\end{equation}
In the limit of small extensions, $\left\langle \Delta z\right\rangle /L\ll1$%
, one recovers $F=\allowbreak\frac{3}{2}\frac{k_{B}T\left\langle \Delta
z\right\rangle }{l_{P}L}$ which is the force that one expects for a Gaussian
coil perturbed by weak forces~\cite{DeGennes}. For large forces one
asymptotically recovers Eq.~\ref{ForceExtensionWLC}.

\section{The Loop in 2D}

In the following we consider a DNA chain under tension that contains a
sliding loop. The corresponding shape at zero temperature is that of the
homoclinic loop \cite{footnote4} that is a solution of the Euler-Lagrange
equations. This filament shape that was already considered by Euler \cite
{Nizette/Goriely} is two-dimensional. For any given finite tension $F$ the
homoclinic loop turns out to be stable for arbitrarily large \textit{in-plane%
} perturbations. Indeed the 2D homoclinic loop can be considered as a
(static) topological soliton appearing in many contexts of contemporary
physics ranging from Josephson-junctions, dislocations in solids \cite
{DavydovSolitons} to QM tunneling problems \cite{Pathintegral}. In the
current section we study the DNA chain being confined to two dimensions;
only in Section IV we go into the third dimension by allowing also
out-of-plane fluctuations. The problem is then that the loop is
intrinsically unstable (in contrast to a false claim in literature \cite
{williams}) and one has to introduce potentials or constraints necessary for
its stabilization.

\subsection{The partition function}

Consider a looped DNA chain under tension $F$ along the $x$-axis. In this
section the DNA is only allowed to fluctuate in-plane (as it is the case for
a chain adsorbed on a fluid membrane, cf.~Fig.~\ref{LoopCompilation}b). We
neglect the DNA twist degree of freedom that in general -- if not explicitly
constrained -- decouples from the DNA bending energy. To obtain the
force-extension behavior of the loop in 2D we evaluate semiclassically the
partition function $Q_{loop}$ by considering quadratic fluctuations around
the saddle point $\phi_{loop}$ that is here the loop configuration. We
impose that the angles at the extremities of DNA are clamped in an
orientation parallel to the force direction, so that $\phi(-L/2)=0$ and $%
\phi(L/2)=2\pi$. Then the partition function in $2D$ corresponds to the
following quantum probability amplitude expressed in terms of a path
integral:
\begin{equation}
Q_{loop}=\langle0,-L/2\lambda|2\pi,L/2\lambda\rangle=\int_{(0,-L/2\lambda
)}^{(2\pi,L/2\lambda)}\mathcal{D}\left[ \phi\right] e^{-\beta E\left[ \phi%
\right] }  \label{2Dpartitionfunction}
\end{equation}
where the energy can be written:
\begin{equation}
E\left[ \phi\right] =\sqrt{AF}\int\nolimits_{-L/2\lambda}^{L/2\lambda
}\left( \frac{1}{2}\dot{\phi}^{2}-\cos\phi\right) dt  \label{EnergyEulers}
\end{equation}
with the dimensionless contour length $t=s/\lambda$; dots represent from now
on derivatives with respect to $t$. In the spirit of the Kirchhoff kinetic
analogy from Section IIA the bending energy in Eq.~\ref{EnergyEulers}
corresponds to the Lagrangian of a spherical pendulum in the gravitational
field. We now expand $E\left[ \phi\right] $ up to quadratic order around the
minimum configuration $\phi_{loop}$ by introducing a fluctuating field $%
\delta\phi$ such that $\phi=\phi _{loop}+\delta\phi$. The linear term $%
\delta E$ in this expansion vanishes because $\phi_{loop}$ is an extremum
point of $E$ and we have
\begin{equation}
E\left[ \phi_{loop}+\delta\phi\right] =E_{loop}+\delta^{2}E\left[ \phi_{loop}%
\right]
\end{equation}

\subsubsection{The saddle point contribution}

To determine the saddle point configuration we solve the Euler-Lagrange
equations of Eq.~\ref{EnergyEulers} that gives the following nonlinear
equation
\begin{equation}
\overset{\cdot\cdot}{\phi}=\sin\phi  \label{SineGordon}
\end{equation}
which is the time independent Sine-Gordon equation well known and studied in
many systems especially in the context of solitons (and their applications
like Josephson junctions, cf. Davydov's book \cite{DavydovSolitons}). Beside
the trivial solution $\phi=0$ that corresponds to the ground state but
cannot describe a loop configuration there exist other topological solutions
of Eq.~\ref{SineGordon} that are appropriately called solitons or kinks. Now
Eq.~\ref{SineGordon} can be integrated twice to obtain
\begin{equation}
\left( t-t_{0}\right) =\int_{\phi\left( t_{0}\right) }^{\phi\left( t\right) }%
\frac{d\phi^{\prime}}{\sqrt{2\left( C-\cos \phi^{\prime}\right) }}
\label{EulerLagrangeThetaIntegrated}
\end{equation}
with an integration constant $C$. The general solution of Eq.~\ref
{SineGordon} with arbitrary $C$ leads to elliptic functions. With the
condition $t_{0}=0$ and $\phi\left( 0\right) =\pi$ the solution reads
\begin{align}
\cos\phi_{loop}\left( t\right) & =2\mathrm{sn}^{2}\left( \frac{t}{\sqrt{m}}%
|m\right) -1  \label{ElliticLoop} \\
\phi_{loop}\left( t\right) & =\pi+2\mathrm{am}\left( \frac{t}{\sqrt{m}}%
|m\right)  \label{PHIloop}
\end{align}
with $\mathrm{sn}$ and $\mathrm{am}$ being the Jacobian elliptic function
with parameter $m$ \cite{Abramowitz Stegun} whose value is related to $C$ in
Eq.~\ref{EulerLagrangeThetaIntegrated} via $m=2/\left( 1+C\right) $. The
parameter $m$ with the range $0\leqslant m\leqslant1$ results from the
clamped boundary conditions and is implicitly given by
\begin{equation}
\sqrt{m}K\left( m\right) =\frac{L}{2\lambda}=\frac{L}{2}\sqrt{F/A}
\label{KmAndLegth}
\end{equation}
with $K\left( m\right)$ denoting the complete elliptic integral of the first
kind \cite{Abramowitz Stegun}. In the Kirchhoff analogy the solution Eq. \ref
{ElliticLoop} describes a revolving pendulum that makes one full turn during
the ''time period'' $L/\lambda$.

The ''classical'' ($T=0$) bending energy of the loop as an implicit function
of the force is then given by
\begin{equation}
\beta E\left[ \phi_{loop}\right] =4\frac{l_{P}}{L}K\left( m\right) \left[
K(m)\left( m-2\right) +4E(m)\right]  \label{E2Dloop}
\end{equation}
where $E\left( m\right) $ is the complete elliptic integral of the second
kind \cite{Abramowitz Stegun}. We compute now the contribution of the
quadratic fluctuations around this looped saddle point to the partition
function.

\subsubsection{The fluctuation contributions}

In the semi-classical approximation \cite{Pathintegral} the partition
function Eq.~\ref{2Dpartitionfunction} can be written as a product of an
energetic contribution and a quadratic path integral over the fluctuating
fields $\delta \phi $ satisfying the Dirichlet boundary conditions $\delta
\phi (-\tfrac{L}{2\lambda })=\delta \phi (-\tfrac{L}{2\lambda })=0$ :
\begin{equation}
Q_{loop}=e^{-\beta E_{loop}}Q_{loop}^{fluct}
\end{equation}
with the partition function corresponding to the quadratic fluctuation
contributions given by
\begin{equation}
Q_{loop}^{fluct}=\int \mathcal{D}\left[ \delta \phi \right] e^{-\tfrac{\beta
\sqrt{AF}}{2}\int_{-L/2\lambda }^{L/2\lambda }\delta \phi \mathbf{\hat{T}}%
\delta \phi dt}=\sqrt{\frac{\beta \sqrt{AF}}{2\pi D(-\frac{L}{2\lambda },%
\frac{L}{2\lambda })}}  \label{2DloopPathIntegral}
\end{equation}
Here $D(-\frac{L}{2\lambda },\frac{L}{2\lambda })$ is the determinant
associated to the quadratic fluctuation operator $\mathbf{\hat{T}}$ that
reads
\begin{equation}
\mathbf{\hat{T}}=\left( \mathbf{-}\frac{\partial ^{2}}{\partial t^{2}}+2%
\mathrm{sn}^{2}\left( \frac{t}{\sqrt{m}}|m\right) -1\right)
\label{TElliptic}
\end{equation}
The problem of finding the eigenvalues of this operator falls into a class
of ''quasi exactly solvable'' problems and typically appears in quantum
mechanical problems. The corresponding differential equation is called the
Lam\'{e} equation \cite{Arscott}. It admits simple solutions in terms of
polynomials of elliptic functions $\mathrm{sn}$, $\mathrm{cn}$ and $\mathrm{%
dn}$. Its discrete spectrum is known \cite{Arscott} and writes
\begin{align}
\nu _{-1}& =\frac{1}{m}-1\text{ and }f_{-1}\left( t\right) =\mathrm{cn}%
\left( \frac{t}{\sqrt{m}}|m\right)  \notag \\
\nu _{0}& =0\text{ and }f_{0}\left( t\right) =\mathrm{dn}\left( \frac{t}{%
\sqrt{m}}|m\right)  \label{eigenfunctionf0} \\
\nu _{1}& =\frac{1}{m}\text{ and }f_{1}\left( t\right) =\mathrm{sn}\left(
\frac{t}{\sqrt{m}}|m\right)  \notag
\end{align}
One sees immediately that the only eigenfunction satisfying the Dirichlet
condition is $f_{-1}$ so that the smallest eigenvalue of $\mathbf{\hat{T}}$
is $\nu _{-1}$ that we denote in the following by
\begin{equation}
\mu _{0}=\frac{1-m}{m}  \label{MU0}
\end{equation}
Therefore for a molecule of finite length there is no zero mode but $\mu
_{0} $ goes to zero in the limit of infinite length $L$ that in terms of $m$
corresponds to the limit $m\rightarrow 1$. The existence of a vanishing
eigenvalue is the consequence of the translational invariance $t\rightarrow
t+t_{0}$ of the loop solution that formally causes a divergence of Eq.~\ref
{2DloopPathIntegral} \cite{Pathintegral}.

The determinant $D\left( -\frac{L}{2\lambda},\frac{L}{2\lambda}\right) $ can
be computed directly via the method of Gelfand and Yaglom \cite
{GelfandYaglom} that consists of solving an initial value problem on the
interval $\left[ -L/2\lambda,L/2\lambda\right] $. The explicit solution for $%
D\left( \frac {L}{2\lambda},-\frac{L}{2\lambda}\right) $ can be stated in
terms of the ''classical'' solution $\phi_{loop}\left( t\right) $:
\begin{align}
D\left( -\frac{L}{2\lambda},\frac{L}{2\lambda}\right) & =\dot{\phi}%
_{loop}\left( \frac{L}{2\lambda}\right) \dot{\phi}_{loop}\left( - \frac{L}{%
2\lambda}\right) \int_{-L/2\lambda}^{L/2\lambda}\frac{dt}{\left( \dot{\phi}%
_{loop}\left( t\right) \right) ^{2}}  \notag \\
& =2\sqrt{m}E\left( m\right)  \label{DEllipticLoop}
\end{align}
This expression, however, has to be taken with caution since it is incorrect
for large values of $L/\lambda$ missing a factor $\sim e^{-\frac{L}{2}\sqrt{%
\frac{F}{A}}}$ that corresponds to the fluctuation contribution of the
linear part of the DNA. To solve this problem one has to take the
translational invariance of the loop into account. The way to deal
rigorously with a zero mode in the infinite $L$ case is well known \cite
{Pathintegral}. One has to consider the infinite number of degenerate saddle
points resulting from the translational invariance by considering the
collective time coordinate $t_{0}$ as an integration variable instead of the
normal mode $a_{0}$ associated to the zero mode eigenfunction $f_0(t)$. For
a finite chain length we also have to take into account the displacement of
the kink solution but this time only on a finite interval of length $L$. We
compute the corrected determinant $D_{corr}$ by removing the would-be-zero
mode from the determinant and by considering explicitly the finiteness of
the space that this mode can populate:
\begin{equation}
\left( D_{corr}\left( -\frac{L}{2\lambda},\frac{L}{2\lambda}\right) \right)
^{1/2}=\frac{\int_{-\infty}^{\infty}e^{-\frac{\beta\sqrt{AF}}{2}%
\mu_{0}a_{0}^{2}}da_{0}}{\int_{-\alpha}^{\alpha}e^{-\frac{\beta\sqrt{AF}}{2}%
\mu_{0}a_{0}^{2}}da_{0}}\left( D\left( -\frac{L}{2\lambda},\frac
{L}{2\lambda}\right) \right) ^{1/2}
\end{equation}
where the boundary $\alpha$ must be determined by computing the Jacobian
defined by $\delta{a_{0}}=J^{-1}\left( m\right) \delta{t_{0}}$. To do this
consider a small translation of the loop which is then given by $\delta
\phi_{loop}=\dot{\phi}_{loop}\delta{t_{0}}$. Because the same translation
can be done by the eigenfunction $f_{0}(t)$ we have $\delta
\phi_{loop}=f_{0}(t)\delta{a_{0}}$. It is easy to check that the
normalized-to-one eigenfunction associated to the zero mode is of the form $%
f_{0}(t)=\left( \int_{-L/2\lambda}^{L/2\lambda}dt\dot{\phi}%
_{loop}^{2}\right) ^{-1/2}\dot{\phi}_{loop}(t)$ from which follows that the
Jacobian is simply equal to the normalized factor
\begin{equation}
J^{-1}\left( m\right) =\left( \int_{-L/2\lambda}^{L/2\lambda}dt\dot{\phi}%
_{loop}^{2}\right) ^{1/2}=\left( \frac{8E(m)}{\sqrt{m}}\right) ^{1/2}
\label{Jacobtrans}
\end{equation}
>From this relation we deduce $\alpha=\frac{L}{2\lambda}\left( \frac {8E(m)}{%
\sqrt{m}}\right) ^{1/2}$ and finally obtain the partition function:
\begin{equation}
Q_{loop}=\left( \frac{\beta\sqrt{AF}}{4\pi\sqrt{m}E\left( m\right) }\right)
^{1/2}\mathrm{erf}\left( \sqrt{\frac{\beta\sqrt{AF}\mu_{0}E(m)}{\sqrt{m}}}%
\frac{L}{\lambda}\right) e^{-2\frac{\beta\sqrt{AF}}{\sqrt{m}}\left[
K(m)\left( m-2\right) +4E(m)\right] }  \label{QloopCorr}
\end{equation}
Using the relations \ref{KmAndLegth} and \ref{MU0} we can rewrite this
expression fully in terms of $m$:
\begin{equation}
Q_{loop}\left( m\right) =\left( \tfrac{l_{P}}{2\pi L}\tfrac{K\left( m\right)
}{E\left( m\right) }\right) ^{\frac{1}{2}}\mathrm{erf}\left( 2\left( 2\frac{%
l_{P}}{L}\right) ^{\frac{1}{2}}K^{\frac{3}{2}}\left( m\right) E^{\frac{1}{2}%
}\left( m\right) \left( 1-m\right) ^{\frac{1}{2}}\right) e^{-4\frac{l_{P}}{L}%
K\left( m\right) \left[ K(m)\left( m-2\right) +4E(m)\right] }  \label{Qloopm}
\end{equation}
Note that the $\mathrm{erf}$-function only differs significantly from unity
for values of $m\approx1$ which corresponds to the long DNA limit $%
L/\lambda\gg1$, i.e., the correction given by the entropic contribution of
the loop is significant only in this limit. This is important because our
computation of the Jacobian is strictly valid only is this limiting case:
The eigenfunction $f_{0}(t)$ associated with the zero eigenvalue $\nu_0$
(and responsible for the translation of the loop) satisfies the Dirichlet
condition only in the limit of an infinitely long chain. This is why the
zero mode (and hence $f_{0}(t)$) are excluded from the determinant for
finite chains. Physically the boundary condition used in our computation
implies that for a finite chain the shift of the loop induces an elastic
deformation that costs energy. Only in the long-chain-limit the loop can
move freely. The constraint Eq.~\ref{KmAndLegth} can be solved for $%
m\approx1 $ giving $m\approx 1-16e^{- \frac{L}{\lambda}}$ showing that the
smallest eigenvalue $\mu_{0}$, Eq.~\ref{MU0}, reads $\mu_{0}=16e^{-\frac{L}{%
\lambda}}$. So indeed -- as intuitively expected -- this eigenvalue becomes
asymptotically zero for $L/\lambda\rightarrow\infty$ (i.e. $m\rightarrow1$).

\subsection{The force-extension relation}

The force-extension relation of the looped chain in 2D follows from the free
energy $G_{loop}=-\beta^{-1}\ln\left(Q_{loop}\right)$ via $\left\langle
\Delta x\right\rangle =-\partial G_{loop}/\partial F$ with $Q_{loop}$ given
by Eq.~\ref{Qloopm}. Due to the structure of $Q_{loop}$ the mean extension
is a sum of three terms: a contribution from the bending energy, a second
one from fluctuations around the loop configuration and a third from the
error function, i.e.
\begin{equation}
\left\langle \Delta x\right\rangle =\left\langle \Delta_{\text{E}
}x\right\rangle +\left\langle \Delta_{\parallel}x\right\rangle +\left\langle
\Delta_{\text{err}}x\right\rangle  \label{ForceExtension2D}
\end{equation}
The saddle point contribution to the mean extension is given by
\begin{equation}
\left\langle \Delta_{\text{E}}x\right\rangle =\frac{L}{m}[\frac
{(2-m)K(m)-2E(m)}{K(m)}]  \label{deltaener}
\end{equation}
The contribution resulting from fluctuations around the loop configuration
(determinant) is given by:
\begin{equation}
\left\langle \Delta_{\parallel}x\right\rangle =\frac{L^{2}}{16l_{p}m}[\frac
{E^{2}(m)+(1-m)K^{2}(m)-2(1-m)E(m)K(m)}{(E(m)K(m))^{2}}]  \label{deltaloop}
\end{equation}
Finally, the contribution coming from the error function is:
\begin{equation}
\left\langle \Delta_{\text{err}}x\right\rangle =\frac{%
L(1-m)[3E^{2}(m)-(1-m)K^{2}(m)-2E(m)K(m)]}{2\sqrt{2\pi}m\mathrm{erf}\left(
2\left( 2\frac{l_{P}}{L}\right) ^{\frac{1}{2}}K^{\frac{3}{2}}\left( m\right)
E^{\frac{1}{2}}\left( m\right) \left( 1-m\right) ^{\frac{1}{2}}\right) }%
\frac{e^{-\frac{8l_{p}}{L}(1-m)E(m)K^{3}(m)}}{E(m)\sqrt{\frac{l_{p}}{L}%
(1-m)E(m)K(m)}}  \label{deltaer}
\end{equation}
This allows one to immediately plot force-extension curves for a loop in 2D.
We dispense here with giving such a plot since the curves turn out to be
very close to the corresponding ones in 3D, presented below in Fig.~\ref
{Curves}. Instead we only extract from Eqs.~\ref{deltaener} to \ref{deltaer}
the limiting cases of small and large forces.

\subsubsection{Limit of small values of $m\approx0.$}

This corresponds to a regime of small forces $L/2\lambda\ll1$, valid only
for chains satisfying $l_{p}/L\gg1.$ In this limit the functions $\mathrm{%
sn(x|m)}$ $\sim\sin(x)$ and $\mathrm{am(x|m)\sim x,}$ so that the loop
configuration given by Eq.~\ref{ElliticLoop} corresponds to a circle $%
\phi_{loop}\left( s\right) =\pi+2s\pi/L+O(m)$. The bending energy is then
given by:
\begin{equation}
\beta E_{\text{circle}}=2\pi^{2}\frac{l_{P}}{L}+O(m)  \label{energycircle}
\end{equation}
where we used $K(m),$ $E(m)$ $\approx\pi/2+O(m)$ for $m$ very small.

To determine the force-extension relation we expand the various
contributions in expression Eq.~\ref{deltaener} to the first order in $m$
and replace $m$ by $\frac{L^{2}}{\pi ^{2}}\frac{F}{A}$. We then arrive at
\begin{equation}
\left\langle \Delta _{\text{E}}x\right\rangle \approx \frac{L}{8\pi ^{2}l_{p}%
}\frac{FL^{2}}{k_{B}T}+O(m^{2})
\end{equation}
which shows that the bending energy contribution to the elongation goes to
zero with the force. This is expected as the bending energy is independent
of the force when $m$ goes to zero by virtue of Eq.~\ref{energycircle}.

In the same manner we obtain for the contribution due to the quadratic
fluctuations around the loop (there is no translational invariance in this
case as the linear part of the chain is negligible) the expression
\begin{equation}
\left\langle \Delta _{\parallel }x\right\rangle \approx \frac{L^{2}}{4\pi
^{2}l_{p}}(1+O(m))  \label{loopm0}
\end{equation}
We find here that the thermal fluctuations cause on average an increase of
the end-to-end distance (resulting in a reduction of the loop size). Note
that this is contrary to the stretching of a linear DNA where entropic
effects lead to a shortening of the polymer (cf.~Eq.~\ref{linearfluctu}).

As the loop cannot slide in this context, the contribution of the error
function to the force-extension relation should vanish. Indeed our
computation gives
\begin{equation}
\left\langle \Delta _{err}x\right\rangle \approx -\frac{120}{\pi ^{2}}\frac{%
L^{2}}{l_{p}}e^{-\tfrac{\pi ^{4}l_{p}}{2L}}(1+O(m^{2}))
\end{equation}
which is negligible due to $l_{p}/L\gg 1$.

In conclusion in the regime $\frac{L}{2\lambda}\ll1,$ the force extension
relation is given by:
\begin{equation}
\frac{\left\langle \Delta x\right\rangle }{L}\approx\frac{L}{4\pi^{2}l_{p}}%
(1+\frac{\beta FL}{2})
\end{equation}
i.e., for short looped chains the extension grows linearly with the force.

\subsubsection{Limit $m\approx 1:$ the homoclinic loop case}

In the limit $m\rightarrow1,$ $K(m)$ diverges as $\ln\left( 4/\sqrt
{1-m}\right) $ and $E(m)\approx1.$ By virtue of Eq.~\ref{KmAndLegth} this
corresponds then to the case $L/2\lambda\gg1$ (strong force regime) where
the length of the molecule is very large compared to the loop size of order $%
\lambda$. If one is only interested in the force extension curve one can
directly take the limit $m\rightarrow1$ in Eq.~\ref{ForceExtension2D}. It
is, however, interesting to rederive it via the saddle point approximation
of the path integral in the infinite chain limit because then the strong
analogy between our computation and the semi-classical treatment of the
tunneling of a quantum particle in a double well potential becomes very
transparent.

\begin{figure}[ptb]
\includegraphics*[width=8cm]{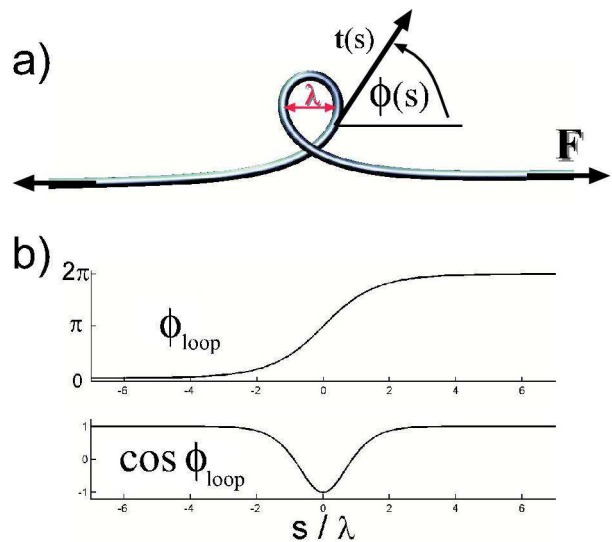}
\caption{a) The definition of the Euler angle $\protect\phi$ and the scale
of the loop. The loop head diameter (red) is approximately given by $\protect%
\lambda=\protect\sqrt{A/F}$ b) The loop solution $\protect\phi_{loop}\left(
s\right) $ as given by Eq.~\ref{Kink2D}.}
\label{HomoclinicLoopTheta}
\end{figure}

In the limit of a very long DNA chain, expressions \ref{KmAndLegth} reduces
to a kink configuration $\phi_{loop}=4\arctan e^{t}$ interpolating between
the two values $\phi(-\infty)=0$ and $\phi (+\infty)=2\pi$, cf.~also Fig.~%
\ref{HomoclinicLoopTheta}. Equation \ref{ElliticLoop} is then given by
\begin{equation}
\cos\phi_{loop}\left( t\right) =1-\frac{2}{\cosh^{2}\left( t\right) } .
\label{Kink2D}
\end{equation}
This saddle point solution is correct only in the infinite chain limit, but
for finite large length the corrections are of order $e^{-L/\lambda}$. This
implies that the bending energy of the kink is then given by
\begin{equation}
E_{loop}=E\left[ \phi_{loop}\right] =-FL+8\sqrt{AF}+O\left( e^{-L/\lambda
}\right) .  \label{EKink}
\end{equation}
The fluctuating quadratic operator in this long chain limit is
\begin{equation}
\mathbf{\hat{T}}=\left( \mathbf{-}\frac{\partial^{2}}{\partial t^{2}}+\left(
1-\frac{2}{\cosh^{2}\left( t\right) }\right) \right)  \label{TOperator}
\end{equation}
This operator is the same as the fluctuating operator obtained by
considering fluctuations around the kink solution that appears in the
semi-classical treatment of a quantum particle in a double well potential.
Having this operator one can compute its set of eigenvalues by solving the
Schr\"odinger equation for a particle moving in a potential of the
Rosen-Morse type (see \cite{Landau-LifshitsQM,Pathintegral}). Then the
partition function is given by
\begin{equation}
Q_{loop}\approx\frac{4}{\pi}\beta LFe^{-\frac{L}{2}\sqrt{\frac{F}{A}}%
}\allowbreak e^{-\beta\left( 8\sqrt{FA}-LF\right) }  \label{Q2Dloop}
\end{equation}
from which we deduce the free energy
\begin{equation}
G_{loop}\approx\frac{L}{2\beta}\sqrt{\frac{F}{A}}+8\sqrt{AF}-LF-\frac{1}{%
\beta}\ln\left( \frac{4}{\pi}\allowbreak\beta LF\right)  \label{GloopFinal}
\end{equation}

We compare now this free energy to that of the straight state. Note that we
cannot use Eq.~\ref{Glinear} since it corresponds to 3D case. Instead we
need the 2D free energy that derives from a partition function that is
evidently given by $Q\left( F,L,T\right) =e^{\beta FL}Q_{1}\left(
F,L,T\right)$. This means that the second and third term of the free energy
expression, Eq.~\ref{Glinear}, have to be divided by 2. Subtracting that
free energy, $G_{0}$, from $G_{loop}$ leads to the free energy difference:
\begin{equation}
\Delta G_{loop-0}=G_{loop}-G_{0}=8\sqrt{AF}+O\left( \frac{1}{\beta}\ln\left(
\frac{LF}{k_{B}T}\frac{l_{P}}{\lambda}\right) \right)  \label{DeltaGKinked0}
\end{equation}
Then we see that the free energy difference $\Delta G_{loop-0}$ is dominated
by the elastic energy part $8\sqrt{AF}$ which is the second term in $%
E_{loop} $, Eq.~\ref{EKink}. The first term $-FL$ is already present in the
straight DNA case and cancels in the difference. Besides that (typically
very large) term there is merely a logarithmic correction. We note that a
weak coupling of the thermal fluctuations to a DNA shape has also been
observed by Odijk~ \cite{Odijk2} for circular DNA rings.

The force-extension curve of a 2D loop is then calculated via $\left\langle
\Delta x\right\rangle =-\partial G_{loop}/\partial F$:
\begin{equation}
\frac{\left\langle \Delta x\right\rangle }{L}=1-\left( \frac{1}{4}+4\frac{
l_{P}}{L}\right) \frac{1}{\sqrt{\beta Fl_{P}}}+\frac{1}{\beta FL} \allowbreak
\label{ForceExtensionKink}
\end{equation}
To understand the origin of each of the various contribution to Eq.~\ref
{ForceExtensionKink} we consider now the limit $m\rightarrow1$ of Eq.~\ref
{ForceExtension2D}. We find for the energetic contribution
\begin{equation}
\left\langle \Delta_{\text{E}}x\right\rangle \underset{m \rightarrow 1}{%
\approx}L\left(1-\frac{4}{L}\sqrt{\frac{A}{F}}\right)
\end{equation}
The contribution of the fluctuation around the saddle point (the
determinant) is given by
\begin{equation}
\left\langle \Delta_{\parallel}x\right\rangle \underset{m \rightarrow 1}{%
\approx}\frac{1}{4\beta F}
\end{equation}
We have already seen in the case of the linear DNA stretching that such a
contribution is negligible in this force regime, cf.~Eq.~\ref{3dxlargeF1}.
That means that in this regime the contribution of the fluctuations to the
force extension relation around the loop are negligible in comparison to the
contribution coming from the elastic energy of the loop.

Now the error function must play an important role as the entropy of the
loop is no more negligible for a very long chain. The contribution from the
error function gives
\begin{equation}
\left\langle \Delta_{\text{err}}x\right\rangle \underset{m \rightarrow 1}{%
\approx}\frac{-L}{4\beta\sqrt{AF}}+\frac{3}{4\beta F}
\end{equation}
The first term in this equation corresponds to the fluctuation of the linear
part (in 2D) of the DNA (compare with the corresponding 3D-term in Eq.~\ref
{3dxlargeF1}) whereas the second term is negligible. Combining the different
contributions we recover equation Eq.~\ref{ForceExtensionKink}. We may drop
the last contribution $k_{B}T/FL$ that is for all practical purposes
negligible.

We can now compare the ''equation of state'' of the looped DNA, Eq.~\ref
{ForceExtensionKink}, with the one for the straight configuration in 2D
given by $\left\langle \Delta x_{0} \right\rangle=-\partial G_{0}/\partial F$%
:
\begin{equation}
\frac{\left\langle \Delta x_{0}\right\rangle }{L}=1-\frac{1}{4}\frac{1}{%
\sqrt{\beta Fl_{P}}}  \label{ForceExtensionStraight}
\end{equation}
Comparing Eq.~\ref{ForceExtensionKink} and Eq.~\ref{ForceExtensionStraight}
we see that both have a leading term proportional to $F^{-1/2}$; only the
prefactor in Eq.~\ref{ForceExtensionKink} is renormalized by a contribution
stemming from the elastic part of the loop free energy.

This implies a fairly simple prediction that is useful for the
interpretation of experimental data: Suppose one performs a single molecule
stretching experiment with a DNA chain that contains a loop. If one does not
know about the presence of the loop one will fit the data by the usual WLC
expression, Eq.~\ref{ForceExtensionStraight}, and is happy that it works
well (at least up to the leading term $F^{-1/2}$). From that fit the total
length of the DNA is recovered correctly (from the asymptotic line on the $%
\Delta x$ axis) but something strange seems to have happened to the
''persistence length'' -- it is smaller than expected. The explanation is
simple: The \textit{apparent persistence length} becomes
\begin{equation}
l_{P}^{app}=\frac{l_{P}}{\left( 1+16\frac{l_{P}}{L}\right) ^{2}}\text{ in 2D}
\label{AppparentLP}
\end{equation}
This formula is similar to Eq.~\ref{Intro1}, announced in the introduction,
with a factor 16 instead of 8 in front of the $l_{P}/L$-term. The difference
comes from the fact that we allow here only fluctuations in 2D. The 3D case
will be studied in chapter IV.

\section{The Homoclinic Loop in 3D}

After having understood the behavior of the homoclinic loop in 2D it seems
that a generalization to the third dimension should be straightforward. But
as we will see there are several traps and some interesting physics on the
way. The first and main problem is the fact that the homoclinic loop is
(unlike in the 2D case) elastically unstable. A simple way to see this is to
take an elastic cable, make a loop in it and to pull on it (without
torsionally constraining the ends). Only if we force the loop to stay in a
plane (for instance, its own weight can perform this task if the cable is
lying on a table provided that we do not pull too strongly) it represents a
\textit{topological excitation} that cannot leave the rod (except at either
of its two ends). So if there is any interesting physics of 3D homoclinic
loops it will have to come through constraints or loop stabilizing
potentials. In the following we mainly consider two stabilizing procedures:
In one case we remove carefully the unstable mode from the partition
function (the loop is then approximately forced to stay in a plane) and in
the second case we evaluate the partition function in the presence of an
explicit self-interaction that stabilizes the loop. It is then shown that
for very long chains the fashion via which the loop is stabilized is
irrelevant with regard to the determination of the force-extension relation.

\subsection{The unstable and rotational zero mode}

We first discuss here the relevance of the right parametrization of the unit
tangent vector. Beside the fact that our parametrization allows us to deal
properly with the measure term, the importance of this choice appears even
more evident when considering the 3D loop. Suppose that we study the
equilibrium property of DNA pulled by a force in the $z$-direction. The
bending energy is
\begin{equation}
E=\int_{-L/2}^{L/2}\left[ \frac{A}{2}\left( \dot{\phi}^{2}\sin ^{2}\theta +%
\dot{\theta}^{2}\right) -F\cos \theta \right] ds
\end{equation}
and the saddle point is now $\phi _{loop}=0$ and $\theta _{loop}$ given by
Eq.~\ref{Kink2D} (with $\phi _{loop}$ replaced by $\theta _{loop}$). Looking
at small variations $\delta \theta $ and $\delta \phi $ around the
homoclinic loop solution we find a \textit{positive definite} second
variation of the energy functional,
\begin{equation}
\delta ^{2}E=\int_{-L/2}^{L/2}\left[ \left( \frac{A}{2}\delta \dot{\theta}%
^{2}+\frac{F}{2}\cos \left( \theta _{loop}\right) \delta \theta ^{2}\right)
+Asin^{2}\left( \theta _{loop}\right) \delta \dot{\phi}^{2}\right] ds
\end{equation}
This is in striking contradiction to the expected elastic instability of the
loop in 3D. The reason is that the coordinate system has a singularity at $%
\theta =0$ where the $\phi $-angle becomes arbitrary. As explained in
section IIB, the way to circumvent the problem is to rotate the force
direction and to put it along the $x$-axis so that the potential energy part
writes now $-F\cos \phi \sin \theta $. In terms of the angles $\phi $ and $%
\vartheta =\theta -\pi /2$ the energy writes now
\begin{equation}
E\left[ \vartheta ,\phi \right] =\sqrt{AF}\int_{-L/2\lambda }^{L/2\lambda
}\left( \frac{1}{2}\left( \dot{\phi}^{2}\cos ^{2}\vartheta +\dot{\vartheta}%
^{2}\right) -\cos \phi \cos \vartheta \right) dt
\label{EThetaPHINewCoordinates}
\end{equation}
with the corresponding Euler-Lagrange equations
\begin{align}
\ddot{\vartheta}& =\cos \phi \sin \vartheta -\dot{\phi}^{2}\cos \vartheta
\sin \vartheta   \notag \\
\ddot{\phi}\cos ^{2}\vartheta -2\dot{\phi}\dot{\vartheta}\cos \vartheta \sin
\vartheta & =\sin \phi \cos \vartheta
\end{align}
We choose in the following the $\vartheta =0$ solution, i.e., we put the
loop into the x-y-plane. This imposes no restriction as we can always rotate
the coordinate system around the $x$-axis to achieve $\vartheta =0$. In this
case we have $\ddot{\phi}=\sin \phi $ which is the same as Eq.~\ref
{SineGordon}, and the saddle point is then given by $\phi _{loop}\left(
t\right) $ , cf.~Eq.~\ref{PHIloop} and $\vartheta _{loop}\left( t\right) =0.$
Then by considering small fluctuations around this saddle point that satisfy
Dirichlet boundary we can again expand Eq.~\ref{EThetaPHINewCoordinates} up
to second order and obtain
\begin{equation}
\beta E\left[ \delta \vartheta ,\phi _{loop}+\delta \phi \right] =\beta
E_{loop}+\tfrac{\beta \sqrt{AF}}{2}\left[ \int_{-L/2\lambda }^{L/2\lambda
}\delta \phi \left( t\right) \mathbf{\hat{T}}_{\parallel }\left( t\right)
\delta \phi \left( t\right) dt+\int_{-L/2\lambda }^{L/2\lambda }\delta
\vartheta \left( t\right) \mathbf{\hat{T}}_{\perp }\left( t\right) \delta
\vartheta \left( t\right) dt\right]   \label{ExpansionRightOperator}
\end{equation}
where the loop energy $E_{loop}$ and the in-plane fluctuation operator $%
\mathbf{\hat{T}}_{\parallel }$ are of course the same as in the 2D case,
cf.~Eq.~\ref{E2Dloop} and Eq.~\ref{TElliptic} respectively. New is in Eq.~%
\ref{ExpansionRightOperator} the out-of-plane fluctuation operator $\mathbf{%
\hat{T}}_{\perp }$ given by
\begin{equation}
\mathbf{\hat{T}}_{\perp }=\mathbf{-}\frac{\partial ^{2}}{\partial t^{2}}+6%
\mathrm{sn}^{2}\left( \frac{t}{\sqrt{m}}|m\right) -\left( \frac{4+m}{m}%
\right) .  \label{Tperpendicular}
\end{equation}
Note that with our choice $\vartheta _{loop}\left( t\right) =0$ the measure
term does not contribute at this level of the approximation. The main
consequence of the quadratic expansion around the saddle point configuration
is that the variables $\vartheta $ and $\phi $ decouple so that the full
partition function $Q_{loop}$ factorizes into the product of the 2D
partition function $Q_{2D}$ (given by Eq.~\ref{Qloopm}) and the partition
function $Q_{\perp }$ accounting for out-of-plane fluctuations:
\begin{equation}
Q_{loop}=Q_{2D}Q_{\perp }
\end{equation}
Although very similar to $\mathbf{\hat{T}}_{\parallel }$ the behavior of the
out-of-plane operator $\mathbf{\hat{T}}_{\perp }$ is fundamentally
different. The discrete spectrum of $\mathbf{\hat{T}}_{\perp }$ consists of
two eigenvalues $\mu _{-1}^{\perp }=-3/m$ and $\mu _{0}^{\perp }=0$, the
first of which is indeed negative \cite{Arscott}.

\begin{figure}[tbp]
\begin{center}
\includegraphics*[width=6cm]{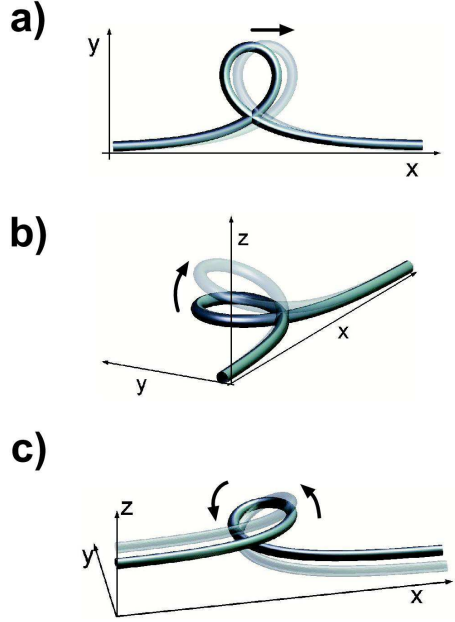}
\end{center}
\caption{The three discrete eigenmodes in action for $m\approx 1$: a) The
translational mode $\protect\phi _{0}=1/\left( \protect\sqrt{2}\cosh
t\right) $ (Eq.~\ref{eigenfunctionf0}), b) The rotational mode $\protect%
\vartheta _{0}=\protect\sqrt{3/2}\sinh t\cosh ^{-2}t $ (Eq.~\ref
{RotationalMode}) and c) The unstable (out of plane tilting) mode $\protect%
\vartheta _{-1}=\protect\sqrt{3/4}\cosh ^{-2}t$ (Eq.~\ref{UnstableModeA}).}
\label{AllDiscreteModes}
\end{figure}

The zero eigenvalue mode of $\mathbf{\hat{T}}_{\perp }$ comes from the
rotational symmetry around the $x$-axis in a similar manner as the
translational invariance of the loop causes a vanishing eigenvalue of $%
\mathbf{\hat{T}}_{\parallel }$. To compute the contribution of the infinite
number of degenerate saddle point related by a rotation around the $x$ -axis
we look at infinitesimal rotational transformations of the loop in 3D. It is
straightforward to show that up to quadratic order a rotation of a kink with
$\vartheta _{loop}=0$ around the $x$-axis by a small angle $\delta
\varepsilon $ corresponds to the following small changes in $\vartheta
_{loop}$ and $\phi _{loop}$:
\begin{align*}
\delta \vartheta _{loop}& \approx \delta \varepsilon \sin \phi _{loop} \\
\delta \phi _{loop}& \approx -\allowbreak \frac{1}{2}\delta \varepsilon
^{2}\sin \phi _{loop}\cos \phi _{loop}=O\left( \delta \varepsilon ^{2}\right)
\end{align*}
We note that in lowest order this rotation leaves $\phi _{loop}$ unaffected,
so that the same rotation $\delta \vartheta _{loop}\approx \delta
\varepsilon \sin \phi _{loop}$ can be done also by the normalized
eigenfunction $\vartheta _{0}(t)$ associated to the zero mode ($\vartheta $
and $\phi $ formally decouple) alone, $\delta \vartheta _{loop}\approx
\delta b_{0}\vartheta _{0}\left( t\right) $ where $b_{0}$ is the normal mode
variable associated to $\vartheta _{0}(t)$. It is easy to check that this
mode normalized to one writes
\begin{equation}
\vartheta _{0}\left( t\right) =\left( \int_{-L/2\lambda }^{L/2\lambda
}dt\sin \phi _{loop}^{2}\right) ^{-1/2}\sin \phi _{loop}
\label{RotationalMode}
\end{equation}
where $\sin \phi _{loop}=\mathrm{cn}\left( \frac{t}{\sqrt{m}}|m\right)
\mathrm{sn}\left( \frac{t}{\sqrt{m}}|m\right) $. Then the Jacobian defined
by $db_{0}=J^{-1}d\varepsilon $ is given by
\begin{equation}
J^{-1}\left( m\right) =\left( \int_{-L/2\lambda }^{L/2\lambda }dt{(\partial
\vartheta _{loop}/\partial \varepsilon )}^{2}\right) ^{1/2}=\left(
\int_{-L/2\lambda }^{L/2\lambda }dt\sin ^{2}\phi _{loop}\right) ^{1/2}
\label{Jacobtheta}
\end{equation}
or explicitly
\begin{equation}
J^{-1}\left( m\right) =2m^{-3/4}\left( \frac{2}{3}\right) ^{1/2}\left( {%
(2-m)E(m)-2(1-m)K(m)}\right) ^{1/2}  \label{Jacobienmfini}
\end{equation}
This relation will be necessary for the computation of the out-of-plane
determinant. In quantum mechanics the ground state wave function has no
node, the first excited state wave function has one node, etc. In our case
the wave function $\vartheta _{0}$ has one node so it cannot be the ground
state and a wave function with no node in the interval considered must
exist. It is obviously the eigenfunction of the unstable mode that is given
by the (unnormalized) expression
\begin{equation}
\vartheta _{-1}\left( t\right) =\mathrm{cn}\left( \frac{t}{\sqrt{m}}%
|m\right) \mathrm{dn}\left( \frac{t}{\sqrt{m}}|m\right)
\label{UnstableModeA}
\end{equation}
with the eigenvalue
\begin{equation}
\mu _{-1}^{\perp }=\frac{-3}{m}  \label{UnstableMode}
\end{equation}
This negative eigenvalue makes the 3D loop mechanically unstable. An
overview of the three discrete eigenmodes is provided in Fig.~\ref
{AllDiscreteModes}.

In the following we will consider three different mechanisms to stabilize
this mode: In the next section we study looped DNA in a strong magnetic
field that breaks the rotational invariance and then we enforce the
cancellation of the unstable mode either by a geometrical constraint
(Section IVB) or by an explicit self interaction (Section IVC), both of
which keep the rotational symmetry.

\subsection{DNA in a strong magnetic field}

A physical situation in which the DNA loop is stabilized is if we switch on
a (very strong) magnetic field along the $z$-axis perpendicular to the force
direction along the $x$-axis, cf.~Fig.~\ref{LoopCompilation}d. The DNA
nucleotides (having $\pi $-electrons) are known to prefer alignment
perpendicular to the field, i.e., DNA exhibits a negative diamagnetic
anisotropy~\cite{Mareth}. The application of a magnetic field $H$ along the
z-axis drives the DNA molecule into a plane parallel to the x-y plane. The
total energy of the DNA writes in this case
\begin{equation}
E\left[ \vartheta ,\phi \right] =\int_{-L/2}^{L/2}\left[\frac{A}{2}\left(
\dot{\phi}^{2}\cos ^{2}\vartheta +\dot{\vartheta}^{2}\right) -F\cos \phi
\cos \vartheta +\frac{\kappa }{2}\sin ^{2}\vartheta \right] ds
\label{EnergyEquatorial}
\end{equation}
The last term gives the coupling between the DNA tangent and the magnetic
field $H$ where $\kappa =-\chi _{a}H^{2}/h$ characterizes the coupling
strength. Here $\chi _{a}$ denotes the (experimentally accessible)
diamagnetic anisotropy of a single DNA basepair \cite{Mareth} and $h=0.34$
nm is the distance between the subsequent DNA basepairs. Note that $\chi
_{a} $ is negative here, i.e., $\kappa >0$ so $\vartheta =0$ is the
preferred rod orientation for large $\kappa $.

Expanding $E\left[ \vartheta ,\phi \right] $ up to second order we obtain
the same expression as Eq.~\ref{ExpansionRightOperator} except that $\mathbf{%
\hat{T}}_{\perp }$ is replaced by a new out-of-plane fluctuation operator $%
\mathbf{\hat{T}}_{\perp }^{\kappa }$
\begin{equation}
\mathbf{\hat{T}}_{\perp }^{\kappa }=\left( \mathbf{-}\frac{\partial ^{2}}{%
\partial t^{2}}+6\mathrm{sn}^{2}\left( \frac{t}{\sqrt{m}}|m\right) -\left(
\frac{4+m}{m}\right) +\frac{\kappa }{F}\right)  \label{TperpendicularCHI}
\end{equation}
The spectrum of Eq.~\ref{TperpendicularCHI} is given by shifting the
spectrum of $\mathbf{\hat{T}}_{\perp }$ by the constant $\kappa /F$ leading
to the eigenvalues $\mu _{s}^{\kappa }=\mu _{s}^{\perp }+\kappa /F$. The
rotational mode is immediately destroyed for any non-zero coupling constant $%
\kappa >0$. More importantly the previously unstable mode $\vartheta _{-1}$
now becomes stable provided that $\kappa /F>3$, i.e., for $\kappa >\kappa
_{crit}=3F$. As the partition function factorizes into the product of the 2D
partition function and an out-of-plane contribution we just need to compute
the determinant associated with the fluctuation operator \ref
{TperpendicularCHI} which is given in Appendix B, Eq.~\ref{DFinalApp}, for
the case $L/2\lambda\gg1$. From this we obtain
\begin{equation}
Q_{\perp}^{\kappa}=\sqrt{\frac{\beta\sqrt{AF}c\left( c+2\right) \left(
c+1\right) }{\pi(c-2)(c-1)}}e^{-c\frac{L}{2\lambda}}  \label{QPerpendicular}
\end{equation}
with $c=\sqrt{1+\kappa/F}$. The free energy $\beta G=-\ln\left(
Q_{2D}Q_{\perp}\right) $ with $Q_{2D}$ given by Eq.~\ref{Q2Dloop} has the
following form
\begin{equation}
\beta G=-\beta FL+8\beta\sqrt{FA}+\frac{\left( 1+\sqrt{1+\frac{\kappa}{F}}%
\right) L\sqrt{F/A}}{2}-\ln\left( \sqrt{\tfrac{2c\left( c+2\right) \left(
c+1\right) }{(c-2)(c-1)}}\tfrac{LA^{1/4}}{\pi^{3/2}\beta^{3/2}}F^{5/4}\right)
\label{Gkinked3D}
\end{equation}
Differentiating this expression with respect to $F$ leads to the
force-extension relation for all forces $F<\kappa/3$. Since this turns out
to be a lengthy expression we give here only the result for the limiting
case $\kappa\gg F$ (and -- as assumed above -- $L/2\lambda\gg1$):
\begin{equation}
\frac{\left\langle \Delta x\right\rangle }{L}=1-\frac{1}{4}\sqrt{\frac{k_{B}T%
}{l_{p}\kappa}}-\sqrt{\frac{k_{B}T}{Fl_{p}}}\left( \frac{1}{4}+4\frac{l_{P}}{%
L}-\frac{1}{8}\left( \frac{F}{\kappa}\right) ^{3/2}+O\left( \frac
{F}{\kappa}\right) ^{5/2}\right) +O\left( \frac{1}{\beta FL}\right)
\label{magnet}
\end{equation}
This expression is similar to the 2D-case, Eq.~\ref{ForceExtensionKink},
which is related to the fact that we assume a strong cost for out-of-plane
fluctuations by setting $\kappa\gg F$. The major contribution from the
out-of-plane fluctuations is the second term on the rhs of Eq.~\ref{magnet}
that describes an effective $F$-independent shortening of the contour
length. The next-order $\kappa$-dependent correction is already by a factor $%
F/\kappa$ smaller and therefore negligible.

Finally, could we experimentally observe the force extension curve derived
above? Unfortunately the coupling parameter $\kappa $ turns out to be too
small for reasonable magnetic fields \cite{footnote6} to be of physical
relevance in practice, i.e. $\kappa \ll 3F$. Nevertheless, the formal
diamagnetic term $\kappa \sin ^{2}\vartheta $ introduced in Eq.~\ref
{EnergyEquatorial} is conceptually useful to understand the (otherwise
unstable) behavior of the DNA loop in 3D. It also turns out to be
technically convenient to use an infinitesimal small ''diamagnetic term'' in
order to break the rotational symmetry of the system for the computation of
the 3D determinant when dealing with the rotational zero mode (cf.~Appendix
A).

\subsection{Force-extension with a geometrical constraint}

In this section we compute the partition function of looped DNA by forcing
the mean tangent of the loop to stay in a plane which is the simplest
stabilizing procedure. This geometrical constraint corresponds to applying
forces at the two chain termini that maintain them in-plane. The constraint
is implemented by the introduction of a delta Dirac distribution  in the
partition function, so that the out-of-plane partition function in the
presence of a external magnetic field becomes
\begin{equation}
Q_{\perp }^{\kappa }\left( \vartheta _{c}\right) =\int \delta \left( \frac{%
\lambda }{L}\int_{-L/2\lambda }^{L/2\lambda }\delta \vartheta dt\right) e^{-%
\tfrac{\beta \sqrt{AF}}{2}\int_{-L/2\lambda }^{L/2\lambda }\delta \vartheta
\mathbf{\hat{T}}_{\perp }^{\kappa }\delta \vartheta dt}\mathcal{D}\left[
\delta \vartheta \right]   \label{QPerpendicularOfD3}
\end{equation}
The formal presence of the external magnetic field is necessary because the
rotational mode $\vartheta _{0}$ corresponds to a zero eigenvalue and causes
a divergence of the partition function $Q_{\perp }^{\kappa =0}$. The problem
results from the fact that a rotation of the kink around the x-axis costs no
energy and consequently the entropic contribution of this state space
direction seems to diverge (within the Gaussian approximation implied by the
saddle point approximation used here). To circumvent this problem we employ
the following trick. Instead of $\mathbf{\hat{T}}_{\perp }$ we use $\mathbf{%
\hat{T}}_{\perp }^{\kappa }$ from Eq.~\ref{TperpendicularCHI} and after
performing all other calculations we let $\kappa \rightarrow 0$ (note that $%
\mathbf{\hat{T}}_{\perp }^{\kappa }|_{\kappa =0}\equiv \mathbf{\hat{T}}%
_{\perp }$). Physically this procedure corresponds to infinitesimally
breaking the rotational symmetry (around the force direction) and restoring
it afterwards in a controlled manner in the limit $\kappa \rightarrow 0$.

Physically it is clear that the main contribution of the mean value of the
tilting angle defined by
\begin{equation}
\left\langle \delta \vartheta \right\rangle =\frac{\lambda }{L}%
\int_{-L/2\lambda }^{L/2\lambda }\delta \vartheta dt
\end{equation}
comes from the unstable mode that induces the large out of plane deviation.
The contribution from the rest of the eigenmodes is small and stable, so
that we can make the following approximation
\begin{equation}
\left\langle \delta \vartheta \right\rangle \approx a_{-1}\left\langle
\vartheta _{-1}\right\rangle =a_{-1}\sqrt{\frac{3}{2}}\frac{m^{1/4}}{%
K(m)\left( (1+m)E(m)-(1-m)K(m)\right) ^{1/2}}  \label{a1only}
\end{equation}
In this way, we approximate the constraint in Eq.~\ref{QPerpendicularOfD3}
by a constraint that fixes the mean out of plane deviation induced by the
unstable mode alone to zero. It means that we relax a bit the constraint in
Eq.~\ref{QPerpendicularOfD3} by allowing the other modes to induce non-zero
mean value of the tilting angle. This contribution however will be small and
limited by the positive spring constants of the stable out of plane modes.

It is then straightforward to rewrite Eq.~\ref{QPerpendicularOfD3} as
follows
\begin{equation}
Q_{\perp }^{\kappa }=\mathrm{i}2\sqrt{|\mu _{-1}|}\left( \sqrt{\frac{\beta
\sqrt{AF}}{2\pi }}\frac{1}{\left\langle \vartheta _{-1}\right\rangle }%
\int_{-\infty }^{\infty }e^{-\tfrac{\beta \sqrt{AF}}{2}\mu
_{-1}a_{-1}^{2}}\delta (a_{-1})da_{-1}\right) \sqrt{\frac{\beta \sqrt{AF}}{%
2\pi D_{\perp }^{\kappa }(-\frac{L}{2\lambda },\frac{L}{2\lambda })}}
\label{Q3Dmodeunstable}
\end{equation}
where -- with $\kappa $ being small -- the determinant is now imaginary.
Note that by removing the unstable mode from the determinant we have taken
into account that the Gaussian integral of the unstable mode is given by
\begin{equation*}
\int_{-\infty }^{\infty }e^{-\mu _{-1}x^{2}/2}\frac{dx}{\sqrt{2\pi }}\equiv
\frac{\mathrm{i}}{2\sqrt{|\mu _{-1}|}}
\end{equation*}
and not by $\mathrm{i/}\sqrt{|\mu _{-1}|}$ as a naive analytical
continuation would suggest \cite{Pathintegral,Langer}. Expression Eq.~\ref
{Q3Dmodeunstable} shows that one can not simply remove the unstable mode
from the determinant but one has to replace it carefully by introducing a
correct constraint expression in the partition function. For instance,
cancelling simply the unstable mode would introduce non physical divergences
in the limit of very small forces.

We are only interested here in the limit of zero magnetic field. In order to
restore the rotational invariance we have to deal with the rotational zero
mode by dividing out the would-be-zero mode and replacing it by the real
physical space it populates. Therefore we have to compute the $\kappa $%
-independent partition function $Q_{\perp }$ defined by
\begin{equation}
Q_{\perp }\underset{\kappa \rightarrow 0}{=}\sqrt{\mu _{0}(\kappa )}\left(
J^{-1}\left( m\right) \int_{0}^{2\pi }\sqrt{\tfrac{\beta \sqrt{AF}}{2\pi }}%
d\epsilon \right) Q_{\perp }^{\kappa }  \label{QQQ}
\end{equation}
where the Jacobian is given by Eq.~\ref{Jacobienmfini}.

The out-of-plane determinant can be deduced from the Gelfand-Yaglom method
which specifies that the determinant can be obtained from the solution of
the following generalized second order Lam\'{e} equation:
\begin{equation}
\left( \mathbf{\hat{T}}_{\perp }+\frac{\kappa }{F}\right) y(t)=0
\label{lamegeneralk}
\end{equation}
The determinant is then given by the relation $D_{\perp }^{\kappa
}(-L/2\lambda ,L/2\lambda )=y(L/2\lambda )$ valid when the conditions $%
y(-L/2\lambda )=0$ and $\dot{y}(-L/2\lambda )=1$ are satisfied \cite
{GelfandYaglom}. A detailed determination of the solution of Eq.~\ref
{lamegeneralk} is provided in Appendix A. In the small $\kappa $ limit the
out-of-plane determinant admits the following expansion (cf.~Eq.~\ref
{detsmallkappa})
\begin{equation}
D_{\perp }^{\kappa }\left( -\frac{L}{2\lambda },\frac{L}{2\lambda }\right) =-%
\frac{\kappa }{F}\frac{2}{3\sqrt{m}(1-m)}\left( (2-m)E(m)-2(1-m)K(m)\right)
\end{equation}
which is negative as it should be because of the presence of the unstable
mode. Combining the different contributions in Eq.~\ref{QQQ} we arrive at
the following expression for the out-of-plane partition function:
\begin{equation}
Q_{\perp }=\frac{2\sqrt{2}}{\sqrt{\pi }}\left( \frac{l_{p}}{L}\right)
^{3/2}K(m)^{5/2}\left( 1-m\right) ^{1/2}\left( \frac{(1+m)E(m)-(1-m)K(m)}{m}%
\right) ^{1/2}  \label{Q3Drenor}
\end{equation}
The complete semi-classical partition function of the looped DNA in 3D is
then given by $Q_{loop}=Q_{2D}Q_{\perp }$ with $Q_{2D}$ given by Eq.~\ref
{Qloopm}.

The mean end-to-end distance has now four contributions
\begin{equation}
\left\langle \Delta x\right\rangle =\left\langle \Delta _{\text{E}
}x\right\rangle +\left\langle \Delta _{\parallel}x\right\rangle
+\left\langle \Delta _{\text{err}}x\right\rangle +\left\langle \Delta
_{\perp }x\right\rangle  \label{deltaplot}
\end{equation}
where the first three expressions are given by Eq.~\ref{deltaener} to \ref
{deltaer} and the out-of-plane contribution obeys
\begin{equation}
\left\langle \Delta _{\perp }x\right\rangle =\frac{L^{2}}{%
16l_{p}mK^{2}(m)E(m)}\left( \frac{%
5(1+m)E^{2}(m)-2(6-3m-m^{2})K(m)E(m)+(7-12m+5m^{2})K^{2}}{%
(1+m)E(m)-(1-m)K(m) }\right)  \label{delatperpx}
\end{equation}

\begin{figure}[tbp]
\begin{center}
\includegraphics*[width=9cm]{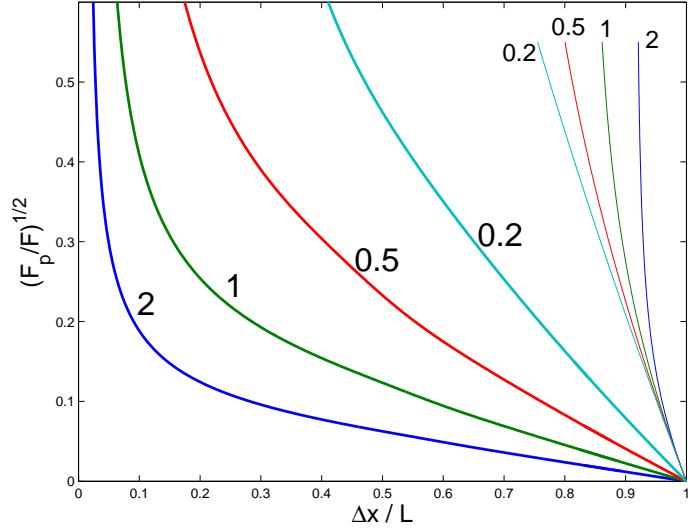}
\end{center}
\caption{Force-extension curve of a DNA chain under tension with a sliding
loop (thick lines, Eq.~\ref{deltaplot}) and without a loop (thin lines, Eq.~%
\ref{3dx1}) for different ratios $l_{p}/L$ as denoted by the numbers close
to the curves. Specifically we plot here $\left(F_p/F\right)^{1/2}$ with $%
F_p=k_B T/l_p$ against $\Delta x/L$. In this representation the curves for
loop-free chains collapse in the limit of large forces, cf. Eq.~\ref
{ExtensionForce}.}
\label{Curves}
\end{figure}

With the complete analytical expression at hand it is straightforward to
compute force-extension curves, some examples for different ratios $L/l_p$
can be found in Fig.~\ref{Curves}. The curves show clearly different scaling
behavior for low and strong forces. In the small $m$ limit $(L/\lambda \ll1)$
-- corresponding to the case $L<l_{p}$ -- Eq.~\ref{delatperpx} has the
following expansion
\begin{equation}
\left\langle \Delta _{\perp }x\right\rangle \approx \frac{L^{2}}{16\pi
^{2}l_{p}}\left( 1+O(m)\right)
\end{equation}
that has the same scaling as the in-plane fluctuation contribution Eq.~\ref
{loopm0}.

In the limit ${L/\lambda \rightarrow \infty }$ or $m\rightarrow 1$ the
out-of-plane partition function, Eq.~\ref{Q3Drenor}, takes the following
form
\begin{equation}
Q_{\perp }\approx \frac{8\sqrt{2}}{\sqrt{\pi }}\left( \beta \sqrt{F}\right)
^{5/4}l_{p}^{1/4}e^{-L/2\sqrt{\frac{F}{A}}}
\end{equation}
which leads to the force-extension curve
\begin{equation}
\left\langle \Delta _{\perp }x\right\rangle \approx -\frac{L}{4}\sqrt{\frac{%
k_{B}T}{Fl_{P}}}+\frac{5k_{B}T}{4F}
\end{equation}
When adding this result to the contributions stemming from the 2D
computation we finally obtain
\begin{equation}
\left\langle \Delta x\right\rangle =L-\frac{1}{2}\sqrt{\frac{k_{B}T}{Fl_{P}}}%
L-4\sqrt{\frac{k_{B}Tl_{P}}{F}}+\frac{9k_{B}T}{4F}  \label{deltax3D}
\end{equation}

In the following section we compute the force-extension curve in the strong
force regime with a loop stabilized by a self-attractive potential. This
computation allows us to check explicitly that the force-extension relation
is fairly independent of the details of the stabilizing procedure and a much
more physical loop-stabilization leads again to Eq.~\ref{deltax3D}.

\subsection{The DNA Self-Attraction and the Homoclinic Loop}

Now we treat an experimentally relevant case in which a loop is stabilized
in 3D: self-attracting DNA. DNA is known to effectively attract itself in
many solvents despite its strong negative bare charge. Typical situations
inducing DNA\ self-attraction are poor solvents (like alcohol, small neutral
polymers like PEG), the presence of multivalent counterions (like CoHex and
Spermidine) or small cationic proteins acting as linkers between two DNA
surfaces. Indeed it was a single molecule stretching experiment on DNA
condensed with multivalent counterions \cite{toroids} that made us think
about the force response of loops.

How should we deal with the DNA self-interaction? A formal treatment that
first comes to ones mind is to introduce a potential $V\left( \left\|
\underline{x}(s_{1})-\underline{x}(s_{2})\right\| \right) $ acting between
any pair of points $s_{1}$ and $s_{2}$ on the DNA molecule and to write the
total interaction energy in form of a double integral (over $s_{1}$ and $%
s_{2}$) as an additional term in our Hamiltonian. The problem is, however,
that we describe the DNA conformation here by the two spherical angles ($%
\vartheta$ and $\phi$) of its\textit{\ tangent} vector whereas the
self-interaction acts in \textit{real space} (''integrated tangent space'').
This makes such a Hamiltonian virtually intractable and hence we need a
reasonable simplification of the DNA self-attraction.

To this end we make here two simplifying assumptions: (\textit{i}) There is
only \textit{a single discrete} DNA\ self-contact point, given by the
crossing point of the homoclinic loop solution. (\textit{ii}) The
interaction potential $V\left( \left\| \underline{x}(s_{1})-\underline{x}%
(s_{2})\right\| \right) $ is \textit{short-ranged} enough so that the
interaction energy at the crossing becomes independent of the crossing
angle, i.e., other parts of the DNA (apart from the crossing point) do not
interact with each other.

These fairly reasonable assumptions imply that the loop ground state
solution will not be significantly modified by the self-attraction and only
the fluctuations around it will be affected. This means that we can write
down the (linearized) loop energy around the solution $\vartheta=0,\phi=%
\phi_{loop}$ in a way similar to the last section, namely
\begin{equation}
E\left[ \delta\vartheta,\phi_{loop}+\delta\phi\right] = E_{loop}+\tfrac{%
\sqrt{AF}}{2}\int_{-L/2\lambda}^{L/2\lambda}\delta\phi\mathbf{\hat{T} }%
_{\parallel}\delta\phi dt+\tfrac{\sqrt{AF}}{2}\int_{-L/2\lambda}^{L/2%
\lambda} \delta\vartheta\mathbf{\hat{T}}_{\perp}^{\kappa}\delta\vartheta dt
+V\left( D_{c}\left( \delta\vartheta\right) \right)
\label{EThetaPHIWithCotact}
\end{equation}
The last term $V\left( D_{c}\right) $ that we introduced here -- in
accordance with above stated assumptions -- represents the interaction
potential of two overcrossing parts of DNA that have a closest distance $%
D_{c} $. To keep the problem tractable we approximate here the distance $%
D_{c}$ by the \textit{perpendicular distance} of the two crossing DNA parts
at the \textit{equilibrium (mean) crossing point} $t_{c}$ of the homoclinic
loop
\begin{equation}
D_{c}\left( \delta\vartheta\right) \approx\lambda\int_{-t_{c}}^{t_{c}}
\sin\delta\vartheta\left( t\right) dt\approx\lambda\int_{-t_{c}}^{t_{c}
}\delta\vartheta\left( t\right) dt  \label{Dcross}
\end{equation}
The crossing point $t_{c}$ will be given by the (in-plane) projected
self-crossing of the loop. This implies the condition that the integral
(over the interval $\left[ -t_{c},t_{c}\right] $) of the $x$-component of
the loop tangent vanishes, i.e., $\int_{-t_{c}}^{t_{c}}\cos\phi_{loop}\left(
t\right) dt=0$ which leads to the following implicit equation for $t_c$:
\begin{equation}
t_{c}=\sqrt{m}\frac{E(t_{c}/\sqrt{m}|m))}{1-m}  \label{tcm}
\end{equation}

Before we compute further it is interesting to have a short look at $D_{c}$
from Eq.~\ref{Dcross}. Because $D_{c}$ depends only on the out-of-plane
perturbations, $\delta\vartheta$, the in-plane ($\delta\phi$) problem stays
unaffected. Note further that the out-of-plane rotational mode $%
\vartheta_{0} $ (the generator of an infinitesimal rotation) leaves the
distance $D_{c}$ unaffected: formally because $\vartheta_{0}\left( t\right) $
is\ an odd function, physically because rotations leave distances fixed.

Now the partition function resulting from Eq.~\ref{EThetaPHIWithCotact} for
any given $V$ can be written as follows
\begin{equation}
Q^{V}=Q_{2D}Q_{\perp}^{V}  \label{QVTotal}
\end{equation}
Only the out-of-plane partition function $Q_{\perp}^{V}$ is modified by the
presence of the contact potential and is given by
\begin{equation}
Q_{\perp}^{V}=\int_{-\infty}^{\infty}e^{-\beta V\left( \vartheta_{c}\right)
}Q_{\perp}\left( \vartheta_{c}\right) d\vartheta_{c}  \label{QperpV}
\end{equation}
where we introduce the angle $\vartheta_{c}=D_{c}/2t_{c}\lambda$ that
measures the (over the loop) averaged angle deviating from the plane $%
\vartheta=0$. The expression $Q_{\perp}\left( \vartheta_{c}\right) $ denotes
the properly constrained partition function
\begin{equation}
Q_{\perp}\left( \vartheta_{c}\right) =\int\delta\left( \vartheta_{c}-\frac{1%
}{2t_{c}}\int_{-t_{c}}^{t_{c}}\delta\vartheta dt\right) e^{-\tfrac{\beta%
\sqrt{AF}}{2}\int_{-L/2\lambda}^{L/2\lambda}\delta \vartheta\mathbf{\hat{T}%
^{\kappa}}_{\perp}\delta\vartheta dt}\mathcal{D}\left[ \delta\vartheta\right]
\label{QPerpendicularOfD}
\end{equation}

To compute this path integral we replace the $\delta$ function by its
Fourier representation
\begin{equation}
\delta\left( \vartheta_{c}-\frac{1}{2t_{c}}\int_{-t_{c}}^{t_{c}}\delta%
\vartheta dt\right) =\frac{1}{2\pi}\int_{-\infty}^{\infty}e^{ip\left(
\vartheta_{c}-\frac{1}{2t_{c}}\int_{-t_{c}}^{t_{c}}\delta\vartheta dt\right)
}dp  \label{DeltaAsFourier}
\end{equation}
The integral in the exponent is more elegantly written as a scalar product
of $\delta\vartheta$ with a ''boxcar'' function $\Pi\left( t\right) =H\left(
t+t_{c}\right) -H\left( t-t_{c}\right) $ with ($H\left( x\right) =1$ for $%
x>0 $, $H\left( x\right) =1/2$ for $x=0$ and $H\left( x\right) =0$
otherwise):
\begin{equation}
\int_{-t_{c}}^{t_{c}}\delta\vartheta
dt=\int_{-L/2\lambda}^{L/2\lambda}\Pi\left( t\right) \delta\vartheta\left(
t\right) dt=\left\langle \Pi|\delta\vartheta\right\rangle  \label{BoxCar}
\end{equation}
where we introduced the scalar product $\left\langle f|g\right\rangle
=\int_{-L/2\lambda}^{L/2\lambda}f\left( t\right) g\left( t\right) dt$. In
this notation and by virtue of Eqs.~\ref{DeltaAsFourier} and \ref{BoxCar}
the partition function $Q_{\perp}\left( D_{c}\right) $, Eq.~\ref
{QPerpendicularOfD}, can be recast in a more transparent form:
\begin{equation}
Q_{\perp}\left( \vartheta_{c}\right) =\frac{1}{2\pi}\int_{-\infty}^{\infty
}e^{ip\vartheta_{c}}\int e^{-\tfrac{\beta\sqrt{AF}}{2}\left\langle
\delta\vartheta\right| \mathbf{\hat{T}}_{\perp}\left| \delta\vartheta
\right\rangle -i\frac{p}{2t_{c}}\left\langle
\Pi|\delta\vartheta\right\rangle }\mathcal{D}\left[ \delta\vartheta\right] dp
\label{QPerpendicularOfD2}
\end{equation}

We have now to compute the following path integral
\begin{equation}
\widehat{Q}_{\perp}^{\kappa}\left( p\right) =\int e^{-\frac{\beta\sqrt{AF} }{%
2}\left\langle \delta\vartheta\right| \mathbf{\hat{T}}_{\perp}^{\kappa
}\left| \delta\vartheta\right\rangle -i\frac{p}{2t_{c}}\left\langle
\Pi|\delta\vartheta\right\rangle }\mathcal{D}\left[ \delta\vartheta\right]
\label{QPerpOfDSimpleFourier*}
\end{equation}
This path integral can be rewritten in as a Gaussian path integral in the
presence of an external source current $j(p,t)=ip\Pi(t)$ coupled linearly to
the fluctuating field $\delta\vartheta$. We refer to Appendix B for the
computation of this kind of path integral. The important point is that it
can be written in the form
\begin{equation}
\widehat{Q}_{\perp}^{\kappa}\left( p\right) =\sqrt{\frac{\beta\sqrt{AF} }{%
2\pi D_{\perp}^{\kappa}\left( -\frac{L}{2\lambda},\frac{L}{2\lambda }\right)
}}e^{-\beta E\left[ j\right] }  \label{QQQQ}
\end{equation}
where the determinant $D_{\perp}^{\kappa}\left( -\frac{L}{2\lambda},\frac
{L}{2\lambda}\right) $ is independent of the source term. Along similar
lines as in Eq.~\ref{QQQ} we go to the limit $\kappa\rightarrow0$:
\begin{equation}
\widehat{Q}_{\perp}(p)\underset{\kappa\rightarrow0}{=}\sqrt{\mu_{0}(\kappa )}%
\left( J^{-1}(m)\int_{0}^{2\pi}\sqrt{\tfrac {\beta\sqrt{AF}}{2\pi}}d\epsilon
\right) \widehat{Q}_{\perp}^{\kappa}(p)  \label{Qrenormalize}
\end{equation}
where the Jacobian is given by expression Eq.~\ref{Jacobienmfini}.

It is very difficult task to calculate Eq.~\ref{QQQQ} for the case of a
finite chain length, so that we restrict ourself to the limit of very long
chains. In this case the determinant is given by Eq.~\ref{DFinalApp} and the
functional $E[j]$ by Eq.~\ref{limS} that is
\begin{equation}
\beta E[j]=-\frac{3}{32}\frac{\left( 3t_{c}^{2}-10\right) }{\beta t_{c}\sqrt{%
AF}}p^{2}
\end{equation}
The implicit condition on $t_{c}$, Eq.~\ref{tcm}, becomes in this limiting
case
\begin{equation}
t_{c}=2\tanh t_{c}  \label{tc}
\end{equation}
that has $t_{c}\approx 1.915$ as the numeric solution. This corresponds to
the actual loop circumference of $2\times 1.915\lambda $. The Jacobian Eq.~%
\ref{Jacobienmfini} in this limiting case obeys $J^{-1}(m=1)=2\sqrt{2/3}$.
Using the fact that the zero eigenvalue can be written as $\mu _{0}\left(
\kappa \right) =\frac{\kappa }{F}=\left( c-1\right) \left( c+1\right) $ we
can compute the partition function Eq.~\ref{Qrenormalize} in the infinite
long molecule limit for $\kappa \rightarrow 0$ (i.e. $c=1$):
\begin{equation}
\widehat{Q}_{\perp }(p)=i8\left( \frac{l_{P}}{\lambda }\right) e^{-\frac{L}{%
2\lambda }}e^{\frac{3\lambda }{32l_{P}t_{c}}\left( 3t_{c}^{2}-10\right)
p^{2}}  \label{Qren}
\end{equation}
Note that because of the unstable mode $\widehat{Q}_{\perp }$ is imaginary.
Transforming back into real space yields
\begin{equation}
{Q}_{\perp }\left( \vartheta _{c}\right) =\frac{1}{2\pi }\int_{-\infty
}^{\infty }e^{ip\vartheta _{c}}\widehat{Q}_{\perp }\left( p\right) dp=8\sqrt{%
\frac{\Gamma }{\pi }}\left( \frac{l_{P}}{\lambda }\right) ^{3/2}e^{-\frac{L}{%
2\lambda }}\exp \left( \tilde{\Gamma}\frac{l_{P}}{\lambda }\left( \vartheta
_{c}\right) ^{2}\right) \text{ }  \label{QConstrFinal}
\end{equation}
where we introduced the scale-independent (negative) elasticity constant for
the out-of-plane tilting
\begin{equation}
\tilde{\Gamma}=\frac{8t_{c}}{3\left( 3t_{c}^{2}-10\right) }  \label{Gamma}
\end{equation}
Using Eq.~\ref{QperpV} we can deduce for any given (reasonable) potential $%
V\left( z\right) $ the out-of-plane partition function
\begin{equation}
Q_{\perp }^{V}=\frac{8}{\lambda }\sqrt{\frac{\Gamma }{\pi }}\left( \frac{%
l_{P}}{\lambda }\right) ^{3/2}e^{-\frac{L}{2\lambda }}\int_{-\infty
}^{\infty }e^{\Gamma \frac{l_{P}}{\lambda }\left( \frac{z}{\lambda }\right)
^{2}-\beta V\left( z\right) }dz  \label{QperpVfinal}
\end{equation}
where we introduced $\Gamma =\tilde{\Gamma}/4t_{c}^{2}$ and the out-of-plane
distance $z\equiv D_{c}=2t_{c}\lambda \vartheta _{c}$. This result indicates
that the larger the perpendicular distance $D_{c}$ the larger the partition
function. This is intuitively clear as the system without the constraint ''$%
D_{c}=const.$'' is intrinsically unstable tending to increase the distance $%
D_{c}$. We can now deduce the full partition function $Q^{V}$ by combining
Eqs.~\ref{Q2Dloop}, \ref{QVTotal} and \ref{QperpVfinal}
\begin{equation}
Q^{V}=\frac{32\sqrt{\Gamma }Ll_{P}^{5/2}}{\pi ^{3/2}\lambda ^{5/2}}e^{\beta
FL-8\frac{l_{P}}{\lambda }-\frac{L}{\lambda }}\int_{-\infty }^{\infty
}e^{\Gamma \frac{l_{P}}{\lambda }\left( \frac{z}{\lambda }\right) ^{2}-\beta
V\left( z\right) }dz  \label{QVGeneral}
\end{equation}
This expression has to be taken seriously only for sufficiently fast growing
interaction potentials $V\left( z\right) $ for which the integral above
stays finite. Otherwise the system is metastable and the integral diverges.
But even in the case when the bound state, say $z=z_{0}$, is just a local
metastable state the integral above $Q^{V}$ still makes some sense if the $%
V\left( z\right) $ is very deep. In this case the system can be considered
as being in quasi-equilibrium (on some experimentally relevant timescale).
If for instance we approximate $V\left( z\right) $ locally by a quadratic
potential $V\left( z\right) =\frac{1}{2}K\left( z-z_{0}\right) ^{2}$ we
obtain
\begin{equation}
Q^{V}=\frac{32\sqrt{2\Gamma }Ll_{P}^{5/2}}{\pi \lambda ^{3}\sqrt{\beta
\lambda ^{3}K-2\Gamma l_{P}}}e^{\frac{\Gamma \beta Kl_{P}}{\lambda ^{3}\beta
K-2\Gamma l_{P}}x_{0}^{2}}e^{\beta FL-8\frac{l_{P}}{\lambda }-\frac{L}{%
\lambda }}\approx \frac{32\sqrt{2\Gamma }Ll_{P}^{5/2}}{\pi \lambda ^{9/2}%
\sqrt{\beta K}}e^{\frac{\Gamma l_{P}}{\lambda ^{3}}z_{0}^{2}}e^{\beta FL-8%
\frac{l_{P}}{\lambda }-\frac{L}{\lambda }}
\end{equation}
The last expression is valid in the limit of strong localization, i.e., for $%
K\lambda ^{2}\gg \sqrt{AF}$.

Let us finally write down the force-extension relation resulting from the
general expression Eq.~\ref{QVGeneral}. We express the free energy $%
G^{V}=-\beta^{-1}\ln Q^{V}$ in terms of $F$ and $A$ (instead of $\lambda=%
\sqrt{A/F}$):
\begin{equation}
\beta G^{V} =-\beta FL+\left( L+8l_{P}\right) \sqrt{\frac{F}{A}}-\ln\left(
\frac{1}{L}\int_{-\infty}^{\infty}e^{\beta\Gamma A^{-1/2}F^{3/2}z^{2}-\beta
V\left( z\right) }dz\right) -\ln\left( \frac{32\sqrt{\Gamma}L^{2}
\beta^{5/2}A^{1/4} F^{9/4}}{\pi^{3/2}}\right)
\end{equation}
This leads to the force extension relation
\begin{equation}
\frac{\left\langle \Delta z\right\rangle }{L} =1-\frac{1}{2}\left( 1+8\frac{%
l_{P}}{L}\right) \frac{1}{\beta\sqrt{AF}}+ \frac{9}{ 4\beta FL}+\frac{3\Gamma%
}{2}\frac {1}{L\lambda} \frac{\int_{-\infty}^{\infty}z^{2}e^{\beta\Gamma
A^{-1/2}F^{3/2} z^{2}-\beta V\left( z\right) }dz}{\int_{-\infty}^{\infty}e^{%
\beta\Gamma A^{-1/2}F^{3/2} z^{2}-\beta V\left( z\right) }dz}
\label{ForceExtensContact}
\end{equation}
The last term $\sim\left\langle D_{c} ^{2}\right\rangle /\left(
L\lambda\right) $ is negligibly small because $\left\langle
D_{c}^{2}\right\rangle $ scales typically as the squared polymer
cross-section (for a short ranged surface contact interaction). In the most
extreme case the contact distance $D_{c}$ could become comparable to $%
\lambda $ (the loop head size), i.e., $\left\langle D_{c}^{2}\right\rangle
\lesssim\lambda^{2}$ but the latter is still much smaller than $L\lambda$.
That means that (for reasonable parameters of $F$ and $L$) the force
extension relation of a DNA loop with attractive contact interaction will
essentially be independent of the concrete realization of the
self-interaction potential $V\left( x\right) $ and we recover the result of
the previous section Eq.~\ref{deltax3D}. The last term $O\left( \left( \beta
FL\right) ^{-1}\right) $ is always negligible for large forces and therefore
the average extension is to a good approximation given by the first two
terms of Eq.~\ref{ForceExtensContact}:
\begin{equation}
\left\langle \Delta x\right\rangle \approx L- \frac{1}{2}\sqrt{\frac{k_{B} T%
}{F l_{P}}}L -4 \sqrt{\frac{k_{B} T l_{P}}{F}}  \label{DeltaZsimple}
\end{equation}
The second term is the usual ''straight WLC'' fluctuation contribution in
3D, Eq.~\ref{ExtensionForce}, the last term is the force extension signature
of the DNA loop, cf. also Eq.~\ref{ForceExtensionKink}.

This computation shows that force-extension relation is fairly independent
of many details like how we stabilize the loop in 3D. The physical reason
for this simple decomposition has its roots in the fact that the WLC
fluctuations (leading to the second term in Eq.~\ref{DeltaZsimple}) and its
state of deformation (third term in Eq.~\ref{DeltaZsimple}) couple only
negligibly in the large force regime (giving merely rise to weak logarithmic
corrections in the free energy and negligible $O\left( \left( \beta
FL\right) ^{-1}\right) $ corrections in the extension $\left\langle \Delta
z\right\rangle $).

The force-extension relation, Eq.~\ref{DeltaZsimple}, has the same
functional form as the usual WLC expression, Eq.~\ref{ExtensionForce}, but
with an apparent persistence length
\begin{equation}
l_{P}^{app} =\frac{l_{P}}{\left( 1+8\frac{l_{P}}{L}\right) ^{2}}
\end{equation}
which is Eq.~\ref{Intro1} of the introduction. This shows that one has to be
cautious when one probes the stiffness of a stiff chain via a stretching
experiment: if the chain contains a loop then one will infer from the data a
value for the chain stiffness that is too small. This is obviously mainly a
problem in cases when the contour length of the chain is not much larger
than its persistence length. But even for $L/l_{P}=10$ one finds $%
l_{P}^{app}\approx0.31l_{P}$ and for $L/l_{P}=50$ there is still a
remarkable effect, namely $l_{P}^{app}\approx0.74l_{P}$.

\section{Conclusion}

We have calculated the partition function of DNA under tension featuring a
sliding loop via a path integration in the semiclassical limit (i.e., on the
level of a saddle point approximation). This path integral can be mapped
onto the QM harmonic oscillator with a time-dependent frequency. In this
analogy the time-dependence reflects the shape of the DNA chain. As it turns
out the planar ground state solutions (Euler elastica) are always just
''simple'' enough to allow the exact solution of the corresponding path
integral. The special choice of the parametrization of the tangent vector to
the DNA has made the application of the semi-classical approximation
possible as the singular measure term (due to inextensibility constraint)
has been found to be negligible in this case.

Within the semiclassical approximation the equation of state of looped DNA
under tension for very stiff polymers is valid for any value of the applied
force. The force-extension relation has been found to be expressed in terms
of Jacobi elliptic functions and the force-extension curve provides two
different scalings for weak and strong forces. For long DNA chains, the
semiclassical approximation is valid only in the regime of strong
stretching. In this force regime we proved that the elastic response of DNA
is (up to logarithmic corrections) indistinguishable from the response of a
non-looped WLC with the same contour length but a smaller persistence
length. As we demonstrated the entropic fluctuations of the system are only
marginally affected by the DNA shape, i.e., the entropies of the overall
straight and of the looped conformation are essentially the same. What
changes considerably when going to the looped state is the enthalpic part.
It is the latter contribution that causes the apparent renormalization of
the chain stiffness. This remarkable effect suggests that the results of
corresponding micromanipulation experiments have to be interpreted
carefully, especially in the case when the contour length of the chain is on
the order of its persistence length.

The looped DNA chain that we presented here should be considered as a
paradigmatic model case. We believe that in the future this powerful
approach will be applicable to a wide range a problems regarding
semiflexible polymers. In Ref.~\cite{kulic} we already applied this method
to DNA chains bearing deflection defects. Analytical results were obtained
in the large force limit for experimentally interesting situations, e.g. for
DNA with a kink-inducing bound protein and the problem of anchoring
deflections in the AFM stretching of semiflexible polymers. Expressions
relating the force-extension curve to the underlying loop/boundary
deflection geometry were provided and applied to the case of the GalR-loop
complex \cite{GalR}. The theoretical predictions were complemented and
quantitatively confirmed by MD simulations \cite{kulic}. Another non-trivial
application of the semiclassical formalism concerns the buckling of rigid
chains \cite{Buckling}, e.g. of microtubuli \cite{janson}.

\appendix

\section{The out-of-plane determinant}

In this appendix we compute the out-of-plane determinant in the presence of
a small magnetic field $D_{\perp }^{\kappa }(-L/2\lambda ,L/2\lambda )$
associated to the following out-of-plane fluctuation operator (cf.~Eq.~\ref
{Tperpendicular})
\begin{equation}
\mathbf{\hat{T}}_{\perp }^{\kappa }=\mathbf{-}\frac{\partial ^{2}}{\partial
t^{2}}+6\text{\textrm{sn} }^{2}\left( \frac{t}{\sqrt{m}}|m\right) -\left(
\frac{4+m}{m}\right) +\frac{\kappa }{F}
\end{equation}
In order to obtain a usual Lam\'{e} equation we consider the transformation $%
t^{\prime }\rightarrow \frac{t}{\sqrt{m}}$ so that $\mathbf{\hat{T}}_{\perp }
$ can be written as
\begin{equation}
\mathbf{\hat{T}}_{\perp }^{\kappa }=\frac{1}{m}\left( \mathbf{-}\frac{%
\partial ^{2}}{\partial t^{\prime 2}}+6m\mathrm{sn}^{2}\left( t^{\prime
}|m\right) -\left( 4+m\right) +3k\right)
\end{equation}
Here we introduce $3k/m=\kappa /F$ for later convenience. It can be shown
\cite{GelfandYaglom} that the determinant is given by a particular solution
of the following generalized second order Lam\'{e} differential equation
\begin{equation}
\mathbf{\hat{T}}_{\perp }^{\kappa }y(t^{\prime })=0  \label{lamegeneral}
\end{equation}
Specifically, the solution satisfying the following boundary conditions:
\begin{equation}
y(-K(m))=0\text{, \ \ and \ \ }\frac{dy}{dt^{\prime }}|_{-K(m)}=\sqrt{m}
\label{boundary}
\end{equation}
gives the desired result via $D_{\perp }^{\kappa }(-\tfrac{L}{2\lambda },%
\tfrac{L}{2\lambda })=y(K(m))$.

We now solve Eq.~\ref{lamegeneral} with a method suggested in Whittaker and
Watson's book \cite{Whittaker}. We first rewrite Eq.~\ref{lamegeneral} in
the form
\begin{equation}
y^{\prime \prime }-\left( 6m\text{sn}^{2}\left( \left. t\right| m\right)
-\varepsilon \right) y=0  \label{Glame2}
\end{equation}
with $\varepsilon =4+m-3k$, and introduce the periodic variable
\begin{equation}
z=\text{cn}^{2}\left( \left. t\right| m\right)   \label{zt}
\end{equation}
in terms of which Eq.~\ref{Glame2} becomes
\begin{equation}
p\left( z\right) y^{\prime \prime }+q\left( z\right) y^{\prime }+r\left(
z\right) y=0  \label{zz}
\end{equation}
with $p\left( z\right) =-mz^{3}+\left( 2m-1\right) z^{2}+\left( 1-m\right) z$%
. Now consider two linear independent solutions $y_{1}\left( z\right) $ and $%
y_{2}\left( z\right) $ from which we build the function $M\left( z\right)
=y_{1}\left( z\right) y_{2}\left( z\right) $. One can then prove that this
function satisfies the following third order differential equation
\begin{equation}
2p\left( z\right) M^{\prime \prime \prime }\left( z\right) +3p^{\prime
}\left( z\right) M^{\prime \prime }\left( z\right) +\left( p^{\prime \prime
}\left( z\right) +2\left( \varepsilon +6m\left( z-1\right) \right) \right)
M^{\prime }\left( z\right) +6mM\left( z\right) =0
\end{equation}
whose solution is a simple periodic function of the form $M\left( z\right)
=z^{2}+az+b$ with the coefficients
\begin{equation}
a=\frac{4-2m-\varepsilon }{3m}\text{ \ \ and \ \ }b=\frac{\left(
1+m-\varepsilon \right) \left( 4+m-\varepsilon \right) }{9m^{2}}
\end{equation}
Because the Wronskian $W$ of Eq.~\ref{zz} is given by $W=y_{1}\dot{y_{2}}%
-y_{2}\dot{y_{1}}=\frac{C}{\sqrt{z(1-z)(1-mz)}}$ we can deduce that the
solutions of Eq.~\ref{zz} are necessarily of the form
\begin{equation}
y_{1,2}\left( z\right) =\sqrt{M\left( z\right) }\exp \left( \frac{\pm C}{2}%
\int \frac{dz}{\sqrt{p\left( z\right) }M\left( z\right) }\right)
\label{y12}
\end{equation}
with
\begin{equation}
C=\frac{1}{9m^{2}}\left[ \left( 4+m-\varepsilon \right) \left(
1+m-\varepsilon \right) \left( 1+4m-\varepsilon \right) \left( \varepsilon
^{2}-4\left( 1+m\right) \varepsilon +12m\right) \right] ^{1/2}
\label{coeffC}
\end{equation}
Introducing the transformation $z=1-\sin ^{2}\phi $ the integral in Eq.~\ref
{y12} can be rewritten
\begin{equation}
J=2\int \frac{d\phi }{\sqrt{1-m\sin ^{2}\phi }M\left( \sin ^{2}\phi \right) }
\label{Jz}
\end{equation}
With the help of the following fractional decomposition
\begin{equation}
\frac{1}{M\left( \sin ^{2}\phi \right) }=\frac{D_{1}}{1-\frac{\sin ^{2}\phi
}{B_{1}}}+\frac{D_{2}}{1-\frac{\sin ^{2}\phi }{^{B_{2}}}}
\end{equation}
where the coefficients are given by
\begin{equation}
B_{1}=\frac{2+a-\sqrt{a^{2}-4ab}}{2}\text{, \ \ \ \ \ \ }B_{2}=\frac{2+a+%
\sqrt{a^{2}-4ab}}{2}  \label{coeffB}
\end{equation}
and
\begin{equation}
D_{1}=\frac{2+a-\sqrt{a^{2}-4ab}}{2\left( 1+a+b\right) \sqrt{a^{2}-4ab}}%
\text{, \ \ \ \ \ \ }D_{2}=\frac{2+a+\sqrt{a^{2}-4ab}}{2\left( 1+a+b\right)
\sqrt{a^{2}-4ab}}  \label{coeffD}
\end{equation}
Eq.~\ref{Jz} can be written in terms of the Jacobi elliptic function of the
third kind $\Pi \left[ n;\varphi \backslash m\right] $
\begin{equation}
J\left( z\right) =2D_{1}\Pi \left[ \frac{1}{B_{1}};\arcsin \sqrt{1-z}%
\backslash m\right] +2D_{2}\Pi \left[ \frac{1}{B_{2}};\arcsin \sqrt{1-z}%
\backslash m\right]   \label{Jzphi}
\end{equation}
with
\begin{equation}
\Pi \left[ n;\varphi \backslash m\right] =\int^{\varphi }\frac{d\phi }{%
\left( 1-n\sin ^{2}\phi \right) \left( 1-m\sin ^{2}\phi \right) ^{1/2}}
\end{equation}
Then formally the solution of Eq.~\ref{Glame2} can be written
\begin{equation}
y_{1,2}\left( t\right) =\sqrt{M\left( t\right) }\exp \left( \frac{\pm C}{2}%
J\left( t\right) \right)   \label{y12solution}
\end{equation}
At this point we mention that for $\kappa >0,$ $C$ is complex. Then the
solution satisfying the boundary conditions Eq.~\ref{boundary} is given by
the following linear combination of two solutions given in Eq.~\ref{y12}:
\begin{equation}
y\left( t\right) =\frac{-\sqrt{m}\sqrt{M\left( -K\right) }}{\left| C\right| }%
\sqrt{M\left( t\right) }\sin \left( \frac{\left| C\right| }{2}\left[ \frac{%
J\left( -K\right) -J\left( t\right) }{2}\right] \right)
\end{equation}
This solution is valid only the interval $-K\left[ m\right] <t<0$ because of
relation Eq.~\ref{zt} between $z$ and $t$. Note that
\begin{equation}
J\left( -K\left[ m\right] \right) =2D_{1}\Pi \left[ \frac{1}{B_{1}},m\right]
+2D_{2}\Pi \left[ \frac{1}{B_{2}},m\right]
\end{equation}
In order to compute the determinant we need the solution $y\left( t\right) $
for $0<t<K\left[ m\right] $ that is also a linear combination of solution
Eq.~\ref{y12}:
\begin{equation}
y\left( t\right) =-\sqrt{m}\frac{\sqrt{M\left( K\right) }}{\left| C\right| }%
\sqrt{M\left( t\right) }\sin \left( \frac{\left| C\right| }{2}(J\left(
t\right) +J\left( K\left[ m\right] \right) \right)
\end{equation}
Then for $t=K\left[ m\right] $ we deduce the determinant $D_{\perp }^{\kappa
}=y\left( K\left[ m\right] \right) $ so that
\begin{equation}
D_{\perp }^{\kappa }=-\sqrt{m}\frac{M\left( K\left[ m\right] \right) }{|C|}%
\sin \left( \left| C\right| J\left( K\left[ m\right] \right) \right)
\label{Det}
\end{equation}
This expression is valid for any value of $\kappa $ and could be used for
the study of looped DNA in strong magnetic fields. Here instead we are
interested in considering the limit of very small $\kappa $ (or,
equivalently, $k$). For this we consider the expansion of the Jacobi
elliptic function.

Expansion of $\Pi\left[ \frac{1}{B_{1}}\backslash m\right] $:

>From Eq.~\ref{coeffB} we deduce the expansion $B_{1}\approx-\frac{1-m}{m^{2}%
}k+O\left( k^{2}\right)$. $B_1$ being negative we find the following
relation \cite{Abramowitz Stegun}:
\begin{equation}
\Pi\left[ n=\frac{1}{B_{1}},m\right] =\frac{n\left( m-1\right) }{\left(
1-n\right) \left( m-n\right) }\Pi\left[ N,m\right] +\frac{m}{m-n}K\left[ m%
\right]
\end{equation}
with $N=\frac{m-n}{1-n}$. As $m<N<1$ we have also $\Pi\left[ N,m\right] =K%
\left[ m\right] +\frac{\pi}{2}\delta_{2}\left( 1-\Lambda_{0}\left(
\varepsilon\right) \right)$ where $\Lambda_{0}\left( \varepsilon\right) $ is
Heuman's Lambda function with
\begin{equation}
\varepsilon=\arcsin\sqrt{\frac{1-N}{\left( 1-m\right) }}=\arcsin\sqrt {\frac{%
1}{1-n}}\approx\frac{1}{m}\sqrt{\left( 1-m\right) k}
\end{equation}
and
\begin{equation}
\delta_{2}=\sqrt{\frac{N}{\left( 1-N\right) \left( N-m\right) }}=\sqrt{\frac{%
\left( m-n\right) \left( n-1\right) }{n\left( 1-m\right) ^{2}}}\approx\frac{m%
}{\left( 1-m\right) \sqrt{\left( 1-m\right) k}}
\end{equation}
>From the relation $\Lambda_{0}\left( \varepsilon\right) =\frac{2}{\pi}%
\left\{ K\left[ m\right] E\left[ \varepsilon,1-m\right] -\left( K\left[ m%
\right] -E\left[ m\right] \right) F\left[ \varepsilon,1-m\right] \right\}$
and using the two expansions $E\left[ \varepsilon,1-m\right]
\approx\varepsilon $ and $F\left[ \varepsilon,1-m\right] \approx\varepsilon$%
, we deduce
\begin{equation}
\Lambda_{0}\left( \varepsilon\right) \approx\frac{2}{\pi}\varepsilon\text{ }E%
\left[ m\right] \approx\frac{2E\left[ m\right] }{\pi m}\sqrt{\left(
1-m\right) k}
\end{equation}
Now Eqs.~\ref{coeffC} abd \ref{coeffD} we the expansions
\begin{equation}
D_{1}\approx-\frac{m^{2}}{\left( 1-m\right) k}\text{ \ \ \ and \ \ }%
|C|\approx\frac{1}{m}\sqrt{k\left( 1-m\right) }
\end{equation}
which allows to write finally
\begin{equation}
2|C|D_{1}\Pi\left[ \frac{1}{B_{1}},m\right] \approx2\sqrt{\frac{\left(
1-m\right) k}{m}}\left( E\left[ m\right] -K\left[ m\right] \right) -\pi
\label{EJ1}
\end{equation}

Expansion of $\Pi\left[ \frac{1}{B_{2}}\backslash m\right] $:

>From Eq.~\ref{coeffB} we deduce the expansion $B_{2}\approx 1+\frac{k}{m^{2}%
}$. Since $B_{2}>1$ and $m<\frac{1}{B_{2}}<1$ we can use the relation
\begin{equation}
\Pi\left[ n=\frac{1}{B_{2}},m\right] =K\left[ m\right] +\frac{\pi}{2}%
\delta_{2}\left( 1-\Lambda_{0}\left( \varepsilon\right) \right)
\end{equation}
where
\begin{equation}
\varepsilon=\arcsin\sqrt{\frac{1-n}{\left( 1-m\right) }}\approx\frac{1}{m}%
\sqrt{\frac{k}{\left( 1-m\right) }}
\end{equation}
and
\begin{equation}
\delta_{2}=\sqrt{\frac{n}{\left( 1-n\right) \left( n-m\right) }}\approx\frac{%
m}{\sqrt{\left( 1-m\right) k}}
\end{equation}
so that finally
\begin{equation}
\Lambda_{0}\left( \varepsilon\right) \approx\frac{2}{\pi}\varepsilon\text{ }E%
\left[ m\right] \approx\frac{2E\left[ m\right] }{\pi m}\sqrt{\frac
{k}{\left( 1-m\right) }}
\end{equation}
As $D_{2}\approx-m^{2}+O(k)$, we find
\begin{equation}
2\left| C\right| D_{2}\Pi\left[ \frac{1}{B_{2}}\backslash m\right] \approx%
\frac{2}{m}\sqrt{\frac{k}{\left( 1-m\right) }}\left( E\left[ m\right]
-\left( 1-m\right) K\left[ m\right] \right) -\pi  \label{EJ2}
\end{equation}
Collecting the two expressions Eq.~\ref{EJ1} and Eq.~\ref{EJ2} we deduce
that
\begin{equation}
\sin\left( \left| C\right| J\left( K\left[ m\right] \right) \right) \approx%
\frac{2}{m}\sqrt{\frac{k}{\left( 1-m\right) }}\left( \left( 2-m\right) E%
\left[ m\right] -2\left( 1-m\right) K\left[ m\right] -2\pi\right)
\label{sinusdet}
\end{equation}
With this result in hand and with the expansion $M\left( 2K\left[ m\right]
\right) \approx\frac{k}{m^{2}}$ we deduce that the determinant Eq.~\ref{Det}
becomes in the small $k$ limit:
\begin{equation}
D_{\perp}^{\kappa}\approx\frac{2k\sqrt{m}}{m^{2}\left( 1-m\right) }\left(
2\left( 1-m\right) K\left[ m\right] -\left( 2-m\right) E\left[ m\right]
\right)  \label{detsmallk}
\end{equation}
which writes in terms of $\kappa$
\begin{equation}
D_{\perp}^{\kappa}\approx\frac{2\kappa}{3F}\frac{1}{\sqrt{m}\left(
1-m\right) }\left( 2\left( 1-m\right) K\left[ m\right] -\left( 2-m\right) E%
\left[ m\right] \right)  \label{detsmallkappa}
\end{equation}

\section{Computation of the perpendicularly constrained partition function
in the limit of infinite long chain}

Here we evaluate the path-integral given by Eq.~\ref{QPerpOfDSimpleFourier*}%
. It is equivalent to a special realization of the path integral of a QM
harmonic oscillator with a time dependent frequency $\omega\left(
\tau\right) $ and a driving force $j\left( \tau\right) $:
\begin{equation}
I\left[ \omega,j\right] =\int_{\left( x_{0},\tau_{0}\right) }^{\left(
x_{1},\tau_{1}\right) }\mathcal{D}\left[ x\right] e^{\frac{i}{ \hbar }S\left[
j,x\right] }=\int_{\left( x_{0},\tau_{0}\right) }^{\left(
x_{1},\tau_{1}\right) }\mathcal{D}\left[ x\right] {e^{\frac{i}{\hbar }\left(
\int_{\tau_{0}}^{\tau_{1}}\frac{m}{2}\left( \dot{x}^{2}\left( \tau\right)
-\omega^{2}\left( \tau\right) x^{2}\left( \tau\right) \right)
d\tau+\int_{\tau_{0}}^{\tau_{1}}j\left( \tau\right) x\left( \tau\right)
d\tau\right) }}  \label{GeneralPathIntegral}
\end{equation}
The latter can be computed exactly (cf.~Refs.~\cite{Pathintegral})
\begin{equation}
I\left[ \omega,j\right] =\sqrt{\frac{m}{2\pi i\hbar D\left( \tau_{1}
,\tau_{0} \right) }}{e^{\frac{i}{\hbar}S \left[ j,x_{cl}\right] } }
\label{GeneralPathIntegralSol}
\end{equation}
The first factor on the rhs of Eq.~\ref{GeneralPathIntegralSol} represents
the fluctuation contribution. Here $D\left( \tau_{1},\tau_{0}\right) $ is
the functional determinant of the ($j$-independent) operator $\mathbf{\hat{T}
}=d^{2}/d\tau^{2}+\omega^{2}\left( \tau\right) $ normalized by the
free-particle operator $d^{2}/d\tau^{2}$
\begin{equation}
\frac{D\left( \tau_{1},\tau_{0}\right) }{\tau_{1}-\tau_{0}}=\det\left( \frac{%
d^{2}/d\tau^{2}+\omega^{2}\left( \tau\right) }{d^{2}/d\tau^{2}}\right).
\label{FunctionalDet}
\end{equation}
The second term in Eq.~\ref{GeneralPathIntegralSol} involves the classical
action $S\left[ j,x_{cl}\right] $ where the $j$-dependent classical path $%
x_{cl}\left( \tau\right) $ is the solution of the corresponding
Euler-Lagrange equation
\begin{equation}
m\ddot{x}_{cl}\left( \tau\right) +m\omega^{2}\left( \tau\right) x_{cl}\left(
\tau\right) =j\left( \tau\right)  \label{XclWithJ}
\end{equation}
with boundary conditions $x_{cl}\left( \tau_{0/1}\right) =x_{0/1}$. Using
Eq.~\ref{XclWithJ} the classical action can be rewritten as
\begin{equation}
S\left[ j,x_{cl}\right] =\frac{m}{2}x_{cl}\left( \tau\right) \dot{x}
_{cl}\left( \tau\right) |_{\tau_{0}}^{\tau_{1}}+\frac{1}{2}\int_{\tau_{0}
}^{\tau_{1}}j\left( \tau\right) x_{cl}\left( \tau\right) d\tau
\label{SjX_cl}
\end{equation}
Now in our concrete case we have to evaluate
\begin{equation}
\widehat{Q}_{\perp}\left( p\right) =\int_{(0,t_{0})}^{(0,t_{1})} \mathcal{D}%
\left[ \delta\vartheta\right] e^{-\tfrac{\beta\sqrt{AF}}{2}%
\int_{t_{0}}^{t_{1} }\delta\vartheta\mathbf{\hat{T}}_{\perp}^{\kappa}\delta%
\vartheta dt+\int_{t_{0}}^{t_{1}}j\left( t\right) \delta\vartheta dt}
\label{GeneralPathIntegralQp}
\end{equation}
with (asymptotic) boundary conditions at $t_{1/0}=\pm\frac{L}{2\lambda }%
\rightarrow\pm\infty$. The operator $\mathbf{\hat{T}}_{\perp}^{\kappa}$ is
again given by
\begin{equation}
\mathbf{\hat{T}}_{\perp}^{\kappa}=\left( \mathbf{-} \frac{\partial^{2}}{%
\partial t^{2}}-\frac{N\left( N+1\right) }{\cosh^{2}\left( t\right) }%
+c^{2}\right)  \label{TKappaApp}
\end{equation}
with $c=\sqrt{1+\frac{\kappa}{F}}$ and $N=2$. The source term is given by $%
j\left( t\right) =-i\frac{p}{2t_{c}}\left( H\left( t+t_{c}\right) -H\left(
t-t_{c}\right) \right) $. In our case the integral \ref
{GeneralPathIntegralSol} (after ''Wick rotation'' $\tau\rightarrow-it$ and
the replacement $1/\hbar\rightarrow\beta,$ $m\rightarrow\sqrt{AF}$, $%
\omega^{2}\left( t\right) \rightarrow\left( 1-6/\cosh^{2}\left( t\right)
+\kappa/F\right) $ etc.) has the following form
\begin{equation}
\widehat{Q}_{\perp}\left( p\right) =\sqrt{\frac{\beta\sqrt{AF}} {2\pi
D\left( t_{1},t_{0}\right) }} e^{-\beta S\left[ j,x_{cl}\right] }
\label{AppQper_p}
\end{equation}
with the classical action given by the following expression
\begin{equation}
\beta S\left[ j,\delta\vartheta_{cl}\right] =\frac{\beta\sqrt{AF}}{2}
\vartheta_{cl}\left( t \right) \dot\vartheta_{cl}\left( t \right)
|_{-\infty}^{\infty}-\frac{1}{2}\int_{-\infty}^{\infty}j\left( \tau\right)
\vartheta_{cl}\left( \tau\right) d\tau= \frac{p^{2}}{2\beta\sqrt{AF}}
\int_{-t_{c}}^{t_{c}}\int_{-t_{c}}^{t_{c}}G\left( t,t^{\prime}\right)
dtdt^{\prime}  \label{Sclassic}
\end{equation}

We first compute the fluctuation determinant $D\left( -\frac{L}{2\lambda },%
\frac{L}{2\lambda }\right) $ via the Gelfand-Yaglom  method. It states that $%
D\left( -\frac{L}{2\lambda },\frac{L}{2\lambda }\right) =f\left( t\right)
|_{t=L/2\lambda }$ where $f\left( t\right) $ is the solution to $\mathbf{%
\hat{T}}_{\perp }^{\kappa }f\left( t\right) =0$ with initial values $f\left(
-\frac{L}{2\lambda }\right) =0$ and $\dot{f}\left( -\frac{L}{2\lambda }%
\right) =1$. The solution can be written in terms of the two linearly
independent solutions (cf.~\cite{Kamke})
\begin{equation}
f_{\pm }\left( x\right) =\cosh ^{N+1}\left( t\right) \left( \frac{1}{\cosh t}%
\frac{d}{dt}\right) ^{N+1}e^{\pm ct}  \label{f_plus_min}
\end{equation}
For $N=2$ the two independent solutions write
\begin{align}
f_{1}\left( t\right) & =e^{ct}\left( (c-2\tanh t)(c-\tanh t)-\cosh
^{-2}\left( t\right) \right)   \notag \\
f_{2}\left( t\right) & =e^{-ct}\left( (c+2\tanh t)(c+\tanh t)-\cosh
^{-2}\left( t\right) \right)   \label{fSolutions12}
\end{align}
and the general solution is given by $f\left( t\right) =C_{1}f\left(
t\right) _{1}+C_{2}f\left( t\right) _{2}$. The Gelfand-Yaglom initial
conditions in the limit $\frac{L}{2\lambda }\gg 1$ (where we may safely set $%
\tanh \left( \pm \frac{L}{2\lambda }\right) \approx \pm 1$, $\cosh
^{-2}\left( \frac{L}{2\lambda }\right) \approx 0$) determine the
coefficients $C_{1}$ and $C_{2}$:
\begin{align*}
C_{1}& =\allowbreak \frac{1}{2c\left( c+2\right) \left( c+1\right) }e^{c%
\frac{L}{2\lambda }} \\
C_{2}& =-\frac{1}{2c\left( c-2\right) \left( c-1\right) }e^{-c\frac{L}{%
2\lambda }}
\end{align*}
Evaluating $f\left( t\right) $ at the right boundary $t=\frac{L}{2\lambda }$
we obtain for $L/2\lambda \gg 1$
\begin{equation}
D_{\perp }^{\kappa }\left( -\frac{L}{2\lambda },\frac{L}{2\lambda }\right)
\approx \frac{(c-2)(c-1)}{2c\left( c+2\right) \left( c+1\right) }e^{c\frac{L%
}{\lambda }}  \label{DFinalApp}
\end{equation}
Note that unlike for the in-plane operator $Q_{\parallel }$ case (where a
close to 0 eigenmode appears and creates artifacts) here we need not to
renormalize $D_{\perp }^{\kappa }\left( -\frac{L}{2\lambda },\frac{L}{%
2\lambda }\right) $ and $Q_{\perp }^{\kappa }$ as long as the value of $c$
is larger than $2$.

To compute the classical action Eq.~\ref{Sclassic}, consider the
Euler-Lagrange equation which reads here
\begin{equation}
\beta\sqrt{AF}\mathbf{\hat{T}}_{\perp}^{\kappa}\delta\vartheta_{cl}\left(
t\right) =j\left( t\right)  \label{InhogEqn}
\end{equation}
with the boundary conditions $\delta\vartheta_{cl}\left( \pm\infty\right) =0$%
. To solve this inhomogeneous differential equation we construct the Green's
function \cite{Barton} $G\left( t,t^{\prime}\right) $ that is the solution
to
\begin{equation}
\mathbf{\hat{T}}_{\perp}^{\kappa}G\left( t,t^{\prime}\right) =\delta\left(
t-t^{\prime}\right)  \label{Green1}
\end{equation}
with $G\left( t,t^{\prime}\right) =G\left( t^{\prime},t\right) $ and proper
boundary conditions $G\left( \pm\infty,t^{\prime}\right) =0$. The latter
gives the solution to Eq.~\ref{InhogEqn} via the simple convolution
\begin{equation}
\delta\vartheta_{cl}\left( t\right) =\left( \beta\sqrt{AF}\right)
^{-1}\int_{-\infty}^{\infty}G\left( t,t^{\prime}\right) j\left( t^{\prime
}\right) dt  \label{Green2}
\end{equation}
For our Dirichlet boundary conditions the Green's function generally writes
\cite{Barton}
\begin{equation}
G\left( t,t^{\prime}\right) =-\frac{H\left( t^{\prime}-t\right) f_{2}\left(
t^{\prime}\right) f_{1}\left( t\right) +H\left( t-t^{\prime }\right)
f_{1}\left( t^{\prime}\right) f_{2}\left( t\right) } {W}  \label{Green3}
\end{equation}
with $f_{1}$ two $f_{2}$ being two (arbitrary) linearly independent
solutions to the homogeneous equation $\mathbf{\hat{T}}_{\perp}^{\kappa}f=0$
satisfying the (one sided) boundary conditions $f_{1}\left( -\infty\right) $
$=0$ and $f_{2}\left( \infty\right) $ $=0$ respectively. The constant $W$ is
the Wronski determinant of the two solutions, i.e.
\begin{equation}
W=f_{1}\left( t\right) \dot{f}_{2}\left( t\right) -\dot{f}_{1}\left(
t\right) f_{2}\left( t\right) =const.  \label{Wronskian}
\end{equation}
We already know the two solutions, cf.~Eq.~\ref{fSolutions12}. Their
Wronskian \ref{Wronskian} is given after short computation by $W=-2c\left(
c^{2}-2\right) \left( c^{2}-1\right) $. Inserting that and Eq.~\ref
{fSolutions12} into Eq.~\ref{Green3} gives a lengthy expression for $G\left(
t,t^{\prime}\right) $. Fortunately there is no need for writing out
explicitly neither $G\left( t,t^{\prime}\right) $ nor $\delta\vartheta
_{cl}\left( t\right) $ as we are only interested in $S\left[ j,x_{cl} \right]
$ from Eq.~\ref{SjX_cl} (with $x_{cl}=\delta\vartheta_{cl}$). Using Eq.~\ref
{Green2} together with the boundary condition $\delta\vartheta _{cl}\left(
\pm\infty\right) =0$ leads to
\begin{equation}
\beta S\left[ j,\delta\vartheta_{cl}\right] = \frac{p^{2}}{4\beta t_{c}^{2}%
\sqrt{AF}} \int_{-t_{c}}^{t_{c}}\int_{-t_{c}}^{t_{c}}G\left( t,t^{\prime
}\right) dtdt^{\prime}
\end{equation}
Inserting the Green's function, Eq.~\ref{Green3}, and exploiting $%
f_{2}\left( t\right) =f_{1}\left( -t\right) $ we find
\begin{equation}
\beta S\left[ j,\delta\vartheta_{cl}\right] =\frac{\left( \beta\sqrt
{AF}\right) ^{-1}p^{2}}{4t_{c}^{2}c\left( c^{2}-2\right) \left(
c^{2}-1\right) } \int_{-t_{c}}^{t_{c}}f_{1}\left( t^{\prime}\right) \left(
\int_{t^{\prime} }^{t_{c}}f_{1}\left( -t\right) dt\right) dt^{\prime}=\frac{%
\left( \beta\sqrt{AF}\right) ^{-1}p^{2}}{4t_{c}^{2}c\left( c^{2}-2\right)
\left( c^{2}-1\right) } I\left( c,t_{c}\right)  \label{Sj}
\end{equation}
The involved double integral $I\left( c,t_{c}\right) $ depends on the
variable $c=\sqrt{1+\frac{\kappa}{F}}$ and the numerical constant $t_{c}$ in
a complicated manner. But here we only need the case $\kappa\rightarrow0$,
i.e., $c\rightarrow1$. The expansion of the integrand around $c=1$ (to
lowest order) followed by the double integration gives (up to terms on the
order of $\left( c-1\right) ^{2}$)
\begin{equation*}
I\left( c,t_{c}\right) =-6\left( c-1\right) \left( 2t_{c}+3\frac{t_{c} }{%
\cosh^{2}t_{c}}-5\tanh t_{c}\right) = \frac{3}{2}t_{c}\left( 3t_{c}
^{2}-10\right) \left( c-1\right) \approx2.88 \left( c-1 \right)
\end{equation*}
Here we made use of the definition of $t_{c}$, Eq.~\ref{tc}. The limit $%
c\rightarrow1$ (i.e. $\kappa\rightarrow0$) can now be performed safely in
Eq.~\ref{Sj} and the action $S\left[ j,\delta\vartheta_{cl}\right] $ writes
\begin{equation}
\lim_{\kappa\rightarrow0}\beta S\left[ j,\delta\vartheta_{cl}\right]
=-\allowbreak\frac{3}{32}\frac{\left( 3t_{c}^{2}-10\right) }{\beta t_{c}
\sqrt{AF}} p^{2} \approx-\frac{0.05}{\beta\sqrt{AF}}p^{2}  \label{limS}
\end{equation}
This leads finally to
\begin{equation}
\widehat{Q}_{\perp}\left( p\right) =\sqrt{\frac{\beta\sqrt{AF}} {2\pi
D\left( t_{1},t_{0}\right) }} e^{-\beta S\left[ j,x_{cl}\right] } =\sqrt{%
\frac{l_{P}}{\lambda}\frac{c\left( c+2\right) \left( c+1\right) }{%
\pi(c-2)(c-1)}}e^{-c\frac{L}{2\lambda}}e^{\frac{3}{32}\frac{\lambda}{l_{P}
t_{c}} \left( 3t_{c}^{2}-10\right) p^{2}}
\end{equation}


\begin{thebibliography}{99}
\bibitem{footnote4}  The term homoclinic stems from the Kirchhoff analogy
between the loop that we consider here and the homoclinic orbit of a
(mathematical) pendulum that obtained just enough energy to make one full $%
2\pi $ rotation in an infinite time interval.
\bibitem{footnote6}  The diamagnetic anisotropy per length $\chi _{a}/h$ of
a single basepair is around $3\times 10^{-6}pN/T^{2}$~\cite{Mareth}, i.e.,
for even very large magnetic fields of say $H=15T$ we have $\kappa \approx
7\times 10^{-4}pN$. So for DNA this effect is rather small.
\end{thebibliography}
\end{document}